\documentclass[letterpaper, 10 pt, conference]{ieeeconf} 
\IEEEoverridecommandlockouts    
\usepackage{cite}
\usepackage{mathptmx}
\usepackage{amsmath,amssymb,amsfonts,amsthm}
\usepackage{algorithmic}
\usepackage{graphicx}
\usepackage{textcomp}
\usepackage{xcolor}
\def\BibTeX{{\rm B\kern-.05em{\sc i\kern-.025em b}\kern-.08em
    T\kern-.1667em\lower.7ex\hbox{E}\kern-.125emX}}

\usepackage[utf8]{inputenc}
\usepackage{cancel}
\usepackage{multicol}
\usepackage{mathrsfs}
\usepackage{enumerate}
\usepackage{listings}
\usepackage{fancyhdr}
\usepackage{tcolorbox}
\usepackage{wasysym}
\usepackage{tikz}
\usepackage{hyperref}
\usepackage{microtype}
\usepackage{lmodern}
\usepackage{braket}
\tcbuselibrary{breakable}

\newcommand{\N}{\mathbb N}

\newcommand{\R}{\mathbb R}

\newtheorem{theorem}{{Proposition}}[section]
\newtheorem{lemma}[theorem]{{Lemma}}
\newtheorem{definition}[theorem]{{Definition}}
\newtheorem{korollar}[theorem]{{Corollary}}
\newtheorem{bemerkung}[theorem]{{Remark}}

\newenvironment{prooof}
 	{\textit{\textit{Proof.}}
	 }
	 {\hfill\qedsymbol\ \\ \ \\
	 }

\hypersetup{
    colorlinks=true,
    linkcolor=blue,
    citecolor=red,
    filecolor=magenta,
    urlcolor=black
}

\newtcolorbox{cdefbox}{colback=green!0!white,colframe=gray!60!white, sharp corners,breakable}
 \newenvironment{ctheo}
 	{\begin{ctheobox}
	 \begin{theorem}}
	 {\end{theorem}
	 \end{ctheobox}}
 
\newtcolorbox{ctheobox}{colback=red!0!white,colframe=gray!60!white, sharp corners,breakable}
 \newenvironment{clem}
 	{\begin{clembox}
	 \begin{lemma}}
	 {\end{lemma}
	 \end{clembox}}
 
\newtcolorbox{clembox}{left=0mm,right=0mm,top=0mm, bottom=0mm,colback=orange!0!white,colframe=gray!60!white, sharp corners,breakable, before={\vspace{0mm}}}
 \newenvironment{ckor}
 	{\begin{ckorbox}
	 \begin{korollar}}
	 {\end{korollar}
	 \end{ckorbox}}
	 \newtcolorbox{ckorbox}{colback=purple!0!white,colframe=gray!60!white, sharp corners,breakable}

\newtcolorbox{cbembox}{colback=red!0!white,colframe=gray!60!white, sharp corners,breakable}
 
 \newtcolorbox{eqbox}{colback=red!0!white,colframe=gray!60!white, sharp corners}

\PassOptionsToPackage{linktocpage}{hyperref}

\begin{document}

\title{Accelerating Extremum Seeking Convergence\\ by Richardson Extrapolation Methods}

\author{Jan-Henrik Metsch$^{1}$,\! Jonathan Neuhauser$^{2}$,\! Jerome Jouffroy$^{3}$,\! Taous-Meriem Laleg-Kirati$^{4}$,\! Johann Reger$^{5}$
\thanks{The authors gratefully acknowledge support by the German Academic Scholarship Foundation for organizing and funding the \emph{Wissenschaftliches Kolleg} during which this project was started and the anonymous referees for their in-depth review. The forth and fifth author gratefully acknowledge funding from the European Union’s
Horizon 2020 Research and Innovation Programme under grant agreement No 824046.
}%
\thanks{$^{1}$J.-H. Metsch, Department of Mathematics, University of Freiburg, Germany ({\tt\small jan.metsch@math.uni-freiburg.de})}%
\thanks{$^{2}$J. Neuhauser, Institute of Fluid Mechanics, Karlsruhe Institute of
Technology, Germany ({\tt\small jonathan.neuhauser@kit.edu})}%
\thanks{$^{3}$Jerome Jouffroy, Department of Mechanical and Electrical Engineering, University of Southern Denmark, Denmark ({\tt\small jerome@sdu.dk})}%
\thanks{$^{4}$Taous-Meriem Laleg-Kirati, Computer, Electrical and Mathematical Sciences and Engineering Division, King Abdullah University of Science and Technology, Saudi Arabia ({\tt\small taousmeriem.laleg@kaust.edu.sa})}%
\thanks{$^{5}$Johann Reger is with the Control Engineering Group, Technische Universit\"at Ilmenau, P.O.\! Box 10\! 05\! 65, 98684 Ilmenau, Germany ({\tt\small reger@ieee.org})}
\thanks{$^\star$Corresponding author: {\tt\small jan.metsch@math.uni-freiburg.de}}%
}

    \maketitle
    \begin{abstract}
    In this paper, we propose the concept of accelerated convergence that has originally been developed to speed up the convergence of numerical methods for extremum seeking (ES) loops. We demonstrate how the dynamics of ES loops may be analyzed to extract structural information about the generated output of the loop. This information is then used to distil the limit of the loop without having to wait for the system to converge to it.  
\end{abstract}

\section{Introduction}
Extremum seeking is a model-free and robust scheme, originally proposed in 1922 by Leblanc (see \cite{leblanc}), to track an extremal operating point of an apparatus by adaptively shifting the operating point in the direction of greatest increase in some output function. The approach has been widely used in the control of systems with \emph{a priori} unknown dynamics. A classical source for an in-depth reference is e.g. \cite{book}, where a proof of convergence is given. Tracking the extremal operating point is achieved by adding a sinusoidal perturbation to the input signal, comparing its phase to the one in the generated output and adjusting the current input based on the phase difference. This is a robust method of tracking an extremal state, but its convergence is rather slow. There are many approaches to analyzing and increasing the speed of convergence as well as eliminating oscillations around the limit available in the literature.  Robustness of several ES methods in application to robotics are discussed in \cite{Calli2012}. The influence of the loop parameters on the speed as well as the domain of convergence is studied in \cite{Nesic2009}. A method to eliminate oscillations around the limit and achieve asymptotic convergence by decreasing the dithering amplitude over time is presented in \cite{Moura2013}. Faster convergence has also been established in \cite{Malek2016} by the usage of fractional operators. Ref. \cite{Haring2016} achieves enhanced convergence for small amplitude and low frequency perturbations by taking the entire plant parameter signals (instead of only the perturbation-related ones) as well as curvature information of the objective function into account. Quite recently Poveda and Kristi{\'c} have introduced the concept of \emph{`prescribed fixed time'}-ES (see \cite{Poveda2020, poveda2020fixedtime}). They accomplish convergence in a given finite time independent of the initial conditions by employing continuous gradient and Newton flows without a Lipschitz property. \\
\vspace{-.1cm}

In this article, we propose to extract the limit directly from the system dynamics. To achieve this, we conduct an in-depth study of the dynamics governing ES to deduce an asymptotic model for the generated output $y(t)$. We then solve the asymptotic model for its limit in terms of the output $y(t)$. This methodology is a form of Richardson extrapolation; a technique originally developed to speed up the convergence of sequences (see \cite{Richardson}). Similar ideas have found applications in a variety of fields such as perturbative quantum field theory (see e.g. \cite{example1,example2}) or machine learning (see e.g. \cite{ML}). The method is, to the best of our knowledge and exhaustive search through the literature, new and has not been applied in the context of control theory.\\
\vspace{-.1cm}

This paper is structured as follows: First, we discuss preliminaries by giving a short introduction to ES and then present the basic idea of accelerated convergence by discussing an ES loop in its most simple form. Next, we demonstrate how to analyze an ES loop theoretically to apply acceleration concepts. We then proceed with some numerical examples to illustrate the performance of the method and close with an outlook on possible future developments. After the bibliography we present detailed proofs.

\section{Preliminaries}
\subsection{Problem formulation}
We consider a function $f: \R\rightarrow\R,\ f(x)=y$ with a local minimum at $x=L$ that we wish to find (for example to optimize a given objective). Such problems appear naturally in many situations such as tracking the optimal operating point of photovoltaic systems  (see e.g. \cite{Brunton2010}) or controlling the optimal substrate flow in bioreactors (see e.g. \cite{bioreactor}). Similar tasks arise in the backpropagation of neural networks (see e.g. \cite{MLbook}, Chapter 4).\\

ES provides an algorithm that continuously improves an initial guess $x_0$ such that the resulting signal $x(t)$ converges exponentially to a neighbourhood of $L$. Intuitively this is achieved by the law
\begin{equation}\label{updatelaw} 
dx=- f'(x(t)) dt.
\end{equation}
To access the value $f'(x)$, a small oscillation $\epsilon\sin(\omega t)$ is added to $x$ leading to
$$f(x+\epsilon \sin(\omega t))=f(x)+\epsilon f'(x)\sin(\omega t)+\mathcal O(\epsilon^2).$$
Running the output of $f$ through a high-pass filter and multiplying with $\sin(\omega t)$ produces the signal $\varphi(t)=\epsilon \sin^2(\omega t)f'(x)$. 
Replacing the actual gradient $f'(x(t))$ in (\ref{updatelaw}) with $\varphi(t)$ gives the law $dx=-\varphi(t)dt$. A block diagram for this process is shown in Fig. \ref{noiseloop}. Closer analysis (see e.g. Chapter 1 in \cite{book}, Equation (1.9)) of this process suggests the approximate formula
\begin{equation}
    x(t)\approx L+C e^{-\epsilon b T}+\epsilon p(t), \label{eq:prbform_1}
\end{equation}
where $p(t)$ is an oscillating function and $C$ and $b$ are constants. The two error terms `\emph{compete}' with each other in the following sense: For large $\epsilon$ the exponential converges rapidly while the oscillating terms becomes large. For small $\epsilon$ the oscillation get suppressed while the exponential decay becomes slow.\\

This motivates studying the dynamics of the ES scheme described above in-depth to `\emph{resolve}' the `\emph{competing objectives}' in \eqref{eq:prbform_1}. The method we propose in this article is essentially designed to eliminate the exponential decay term in \eqref{eq:prbform_1} which allows for fast convergence for sufficiently small values of $\epsilon$.

\subsection{Accelerated convergence}
We present an easy example of accelerated convergence. A detailed review can be found in \cite{benderreview}. Consider the sequence $S_n:=\sum_{j=1}^ n\frac 1{j^2}$. It is well known that $S_n\rightarrow\frac{\pi^2}6$. The convergence is very slow however as 
\begin{equation}\label{model}
\frac{\pi^2}6-S_n\sim\int_n^ \infty\frac {dt}{t^2}=\frac1n.
\end{equation}
To accelerate the convergence, we first construct an asymptotic model. Motivated by (\ref{model}) it is reasonable to assume (and not too hard to prove) an expansion of the form
\begin{equation}\label{Snexpansion}
S_n=L+\sum_{j=1}^\infty\frac{a_j}{n^ j}.
\end{equation}
Here we abbreviated the limit of $S_n$ as $L:=\frac{\pi^2}6$. A quick calculation shows that 
\begin{equation}\label{SnExtraction}\tilde S_n:=\frac12\left((n+2)^2S_{n+2}-2(n+1)^2 S_{n+1}+n^2 S_n\right)
\end{equation}
satisfies $\tilde S_n=L+\mathcal O\left(\frac 1{n^3}\right)$. Hence, the convergence has been accelerated. Indeed $L=1.64493$, $\tilde S_{10}=1.64481$ while $S_{10}=1.54976$.

\section{Theory}
We show how the concept of accelerated convergence may be applied to ES by studying two distinct loops starting with the easiest one and then demonstrating how a more complex situation may be analyzed. 
For the latter, we need to perform perturbation analysis to extract structural information about the dynamics. We remark that regular dependence of solutions on a perturbation parameter is a standard result and e.g. discussed in \cite{Amann}, Chapter 2, Section 9.  The analysis essentially aims to derive a precise version of \eqref{eq:prbform_1} similar to \eqref{Snexpansion}. Considering shifts in time $t\rightarrow t+T$ we then derive extraction schemes for the limit of the system, similar to \eqref{SnExtraction}.
Finally, we point out that a similar analysis has been performed in \cite{Bender} for the Mathieu equation (see Chapter 11, Section 4). 
\subsection{Basic model}\label{basicmodel}
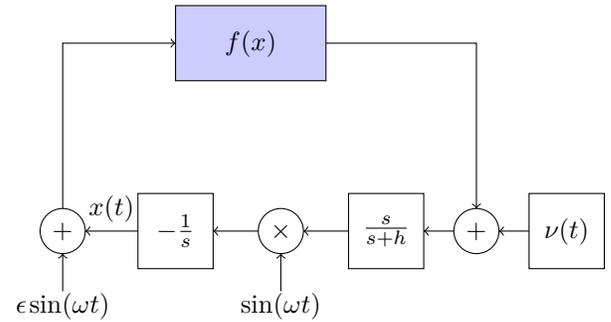
\begin{figure}[h!]
\begin{center}
\begin{tikzpicture}
\draw[fill=blue!20](-1.5,.5) rectangle (0.5,1.5) node[pos=0.5] {$f(x)$};
\draw(1.8,-2) rectangle (.8,-1)node at (1.3,-1.5){$\frac{s}{s+h}$};
\draw (-.1,-1.5) circle (.3)node at (-.1,-1.5){$\times$};
\draw(-1,-2) rectangle (-2,-1) node at (-1.5,-1.5){$-\frac 1s$};
\draw (-3,-1.5) circle (.3) node at (-3,-1.5){$+$};
\draw[->](-3,-2.3)--(-3,-1.8) node at (-3,-2.5) {$\epsilon\sin(\omega t)$};
\draw(4.2,-2) rectangle (3.2,-1) node at (3.7,-1.5){$\nu(t)$};
\draw (2.5,-1.5) circle (.3) node at (2.5,-1.5){$+$};
\draw[->](0.5,1)--(2.5,1)--(2.5,1)--(2.5,-1.2);
\draw[->](.8,-1.5)--(.2,-1.5);
\draw[->](-.4,-1.5)--(-1,-1.5);
\draw[->] (-3,-1.2)--(-3,1)-- (-1.5,1);
\draw[->](-.1,-2.3)--(-.1,-1.8) node at (-.1,-2.5) {$\sin(\omega t)$};
\draw[->] (-2,-1.5)--(-2.7,-1.5) node[pos=0.5, above] {$x(t)$};
\draw[->] (2.2,-1.5)--(1.8,-1.5);
\draw[->] (3.2,-1.5)--(2.8,-1.5);
\end{tikzpicture}
\end{center}
\caption{Extremum seeking loop}\label{noiseloop}
\end{figure}

Let $a,b,L\in\R$ and $f(x):=a+b(x-L)^2$. Initially, we analyze the ES loop depicted in Fig. \ref{noiseloop}.
\ \\
$\nu (t)$ is a noise source, which will be included in the simulations in Section \ref{simulations}. Denoting the high-pass filter by $\mathcal F$, Figure \ref{noiseloop} corresponds to the integral equation \begin{equation}\label{loopeq}
\begin{aligned}
x(t)=x(0)-&\int_0^t\mathcal F\left[f\left(x(\tau)+\epsilon\sin(\omega\tau)\right)\right.\\
&\left.+\nu(t)\right]\sin(\omega\tau)d\tau.
\end{aligned}
\end{equation}

\begin{ctheo}\label{basictheo}
Let $T:=\frac{2\pi}\omega$, $\theta:=e^{-\epsilon bT}$ and $x:[0,\infty)\rightarrow\R$ be a solution to the loop in Fig. \ref{noiseloop} with $\nu\equiv 0$. For any $t\geq 0$, put $x_n:=x(t+nT)$. Then
$$L=\frac{(x_0-x_1)x_2+\theta x_0(x_2-x_1)}{x_0-(1+\theta)x_1+\theta x_2}+\mathcal     O(\epsilon^2).$$
Additionally, putting 
$$g:=\frac{(x_0-x_1)(x_2-x_3)}{(x_1-x_2)(x_0-x_3)}$$
the following extraction law for $\theta$ holds:
$$\theta=\frac{1-g}{2g}-\frac1{2g}\sqrt{-4g^2+(g-1)^2}+\mathcal O(\epsilon^2)$$
\end{ctheo}
\begin{prooof}
Let $y(t):=x(t)-L$. Then $\dot y=\dot x$. Differentiating \eqref{loopeq} and using $n\equiv 0$ gives 
\begin{align}
&\dot y+\epsilon b(1-\cos(2\omega t))y+by^2\sin(\omega t)\nonumber\\
=&-b\epsilon^2\sin(\omega t)^3.\label{basicyeq}
\end{align}
This is a Ricatti equation without a closed-form solution. We consider $\epsilon$ as a perturbative parameter and only study (\ref{basicyeq}) to first order. This justifies dropping the $\epsilon^2$-term in (\ref{basicyeq}) which gives a Bernoulli Equation. Putting
$$x_0(t):=\exp\left[-\epsilon bt+\frac {\epsilon b}{2\omega}\sin(2\omega t)\right]$$
we derive the following formula for its solution $x$ in Appendix \ref{app:eq4}:
\begin{equation}\label{sol}
x(t)=L+\frac{ x_0(t)}{C+b\int_0^t\sin(\omega s)x_0(s)ds}.
\end{equation}
The constant $C$ is related to the initial value $x(0)$. Recalling $\theta=e^{-\epsilon b T}$, it is clear that $x_0(t+T)=\theta x_0(t)$. Let $\varphi(t):=C+b\int_0^t\sin(\omega s)x_0(s)ds$ so that $\dot\varphi(t+T)=\theta\dot\varphi(t)$. Lemma \ref{calclem} in Appendix \ref{app:calclem} implies $\varphi(t)=\tilde C+X(t)$ for a constant $\tilde C$ and a function $X$ satisfying $X(t+T)=\theta X(t)$. This gives the following equations: 
\begin{equation}\label{system1}
\begin{aligned}
x(t)-L&=\frac{x_0(t)}{\tilde C+X(t)}\\
x(t+T)-L&=\theta\frac{x_0(t)}{\tilde C+\theta X(t)}\\
x(t+2T)-L&=\theta^2\frac{x_0(t)}{\tilde C+\theta^2 X(t)}
\end{aligned}
\end{equation}
If we regard $x(t+nT)$ as known parameters, \eqref{system1} can be thought of as a nonlinear system of ordinary equations for $L, \tilde C, X(t)$ and $x_0(t)$. A solution for $L$ then gives a formula of the limit in terms of the values $x_n:=x(t+nT)$. Direct computation shows 
\begin{equation}\label{extract1}
L=\frac{(x_0-x_1)x_2+\theta x_0(x_2-x_1)}{x_0-(1+\theta)x_1+\theta x_2}.
\end{equation}
Equation (\ref{extract1}) uses the data points $x(t)$, $x(t+T)$ and $x(t+2T)$ and fits them onto the solution (\ref{sol}). It eliminates the unknown values $x_0(t)$, $\tilde C$ and $X(t)$ and hence requires three data points. Note however that $\theta=e^{-\epsilon bT}$ features in the extraction law. While $T$ and $\epsilon$ are part of the design of the loop and therefore known, the parameter $b$ is part of the function $f$ and in general not known. By incorporating a fourth data point into the analysis we can eliminate $\theta$ from (\ref{extract1}). Indeed we note that (\ref{extract1}) also holds for $t\rightarrow t+T$ and hence 
\begin{align} 
L=&\frac{(x_0-x_1)x_2+\theta x_0(x_2-x_1)}{x_0-(1+\theta)x_1+\theta x_2}\nonumber\\
=&\frac{(x_1-x_2)x_3+\theta x_1(x_3-x_2)}{x_1-(1+\theta)x_2+\theta x_3}\label{LisLidentity}.
\end{align}
This is a quadratic equation for $\theta$ with two solutions. However, putting
\begin{equation}\label{gdefinition}
g:=\frac{(x_0-x_1)(x_2-x_3)}{(x_1-x_2)(x_0-x_3)}
\end{equation}
we prove in Appendix \ref{thetaeqproof} that
\begin{equation}\label{theaeq}
\theta=\frac{1-g}{2g}-\frac1{2g}\sqrt{-4g^2+(g-1)^2}
\end{equation}
by exploiting $\theta=e^{-\epsilon bT}\in(0,1)$.
\end{prooof}
We have derived an extraction scheme that uses four data points. It first applies (\ref{theaeq}) to find $\theta$ and then uses (\ref{extract1}) to extract the limit $L$.

\subsection{Including a drift}
This Subsection demonstrates how to extend the analysis from Subsection \ref{basicmodel} to other loops by considering an example. We modify the ES loop in Fig. \ref{noiseloop} by taking $f(x,t)=(x-L-q(t))^2$ to be explicitly time dependent. 
We refer to the resulting loop as modified Fig. \ref{noiseloop}. Here $q(t)=q_0e^{-\delta t}$ for a small positive drift parameter $\delta>0$.
\begin{ctheo}\label{drifttheorem}
Let $x$ be any solution of modified Fig. \ref{noiseloop} with $\nu\equiv 0$ and put $z(t):=(x(t)-L-q(t))^{-1}$. Then
$$z(t)=\sum_{j=0}^\infty \left[\delta^je^{-j\delta  t}\sum_{k=0}^{j+1}e^{k\epsilon t} p_{jk}(t)\right]+O(\epsilon^2).$$
where all function $p_{jk}$ are $T$-periodic. 
\end{ctheo}
\begin{prooof}
We put $y:=x-L-q$. Differentiating the analogue of \eqref{loopeq} with time-dependent $f$ and exploiting  and $\dot q(t)=-\delta q(t)$ gives 
\begin{equation}\label{driftode}
    \dot y(t)+2\epsilon\sin(\omega t)^2 +y^2\sin\omega t-\delta q=-\epsilon^2\sin(\omega t)^3
\end{equation}
After dropping $\epsilon^2$ as in the proof of Proposition \ref{basictheo} and letting $z=\frac1y$, we get 
\begin{equation}\label{zODE}
\dot z-2\epsilon \sin^2(\omega t)z+\delta q(t)z^2=\sin(\omega t).
\end{equation}
Equation (\ref{zODE}) is another Riccati equation without closed-form solution. Still we may extract structural properties by perturbation analysis. Proposing $z(t)=\sum_{n\geq 0} z_n(t)\delta^n$ we get the following infinite system of linear ordinary differential equations: For $n=0$:
\begin{equation}\label{z0ODE}
\left\{
    \begin{aligned}
    &\dot z_0-2\epsilon\sin^2(\omega t)z_0=\sin(\omega t)\\
    &z_0(0)=z(0)
    \end{aligned}
    \right.
\end{equation}
For $n\geq 1$:
\begin{equation}\label{znODE}
\left\{
    \begin{aligned}
    &\dot z_n-2\epsilon\sin^2(\omega t)z_n=-q(t)\sum_{j=0}^{n-1}z_jz_{n-1-j}\\
    &z_n(0)=0
    \end{aligned}
    \right.
\end{equation}
Solving for $z_0$ is trivial. Working iteratively, the $n$-th equation is linear in $z_n$ with nonlinearities only in the already known functions $z_k$ with $k\leq n-1$. An inductive argument shows 
\begin{equation}\label{znformula}
z_n(t)=e^{\epsilon t}p^{(n)}_0(t)+\sum_{j=1}^n \sum_{k=0}^{j+1}e^{(k\epsilon-j\delta) t}p^{(n)}_{jk}(t)
\end{equation}
with $T$-periodic functions $p^{(n)}_{*}$ for $n\geq 1$ and $z_0(t)=p^{(0)}_0(t)+e^{\epsilon t}p^{(0)}_1(t)$ with $T$-periodic functions $p^{(0)}_*$. 
Resumming gives the Lemma.
\end{prooof}
To derive an exact extraction scheme from the expansion given in Proposition \ref{drifttheorem}, we would require infinitely many data points to eliminate all terms in the series. For small $\delta$ we may, however, truncate the perturbation series and construct a finite extraction scheme, which we demonstrate in the following Corollary. 

\begin{ckor}
Let $A:=e^{\epsilon T}$ and $x$ be any solution to the loop in modified Fig. \ref{noiseloop} with $\nu\equiv 0$. Put $h(t):=x(t)-q(t)$ and $h_n:=h(t+nT)$. Then
\begin{equation}\label{phil}
L=\frac{h_1h_0-(1+A)h_2h_0+A h_2h_1}{-h_2+(1+A)h_1-A h_0}
\end{equation}
up to an error of order $\mathcal O(\delta)+\mathcal O(\epsilon^2)$.
\end{ckor}
\begin{prooof}
As it is not entirely trivial, we also demonstrate how to derive a $\mathcal O(\delta^2)$-extraction law. 
Let $B:=e^{\-\delta T}$ and  $z_a:=z(t+aT)$. It is readily checked that
\begin{align*}
    0=&z_5-(1+A+B(1+A+A^2)z_4\\
    &+\big(A+B(1+A)A(1+A+A^2)\\
    &\hspace{3cm}+B^2A(1+A+A^2)\big)z_3\\
    &-\big((AB(1+A+A^2)+(A+1)B^2(A+A^2+A^3)\\
    &\hspace{3cm}+A^3B^3\big)z_2\\
    &+(AB^2(A+A^2+A^3)+(A+1)A^3B^3)z_1\\
    &-A^4B^3 z_0.
\end{align*}
Summarizing this as $\sum_{0\leq i\leq 5}\mu_iz_i=0$ and recalling the definition of $z$ we get the implicit extraction law
$$\sum_{i=0}^5\mu_i\prod_{\substack{j=0\\ j\neq i}}^5(x_j-B^jq_0-L)=0.$$
For zero order extraction scheme one argues analogously. Solving the resulting implicit law gives (\ref{phil}).
\end{prooof}
Note that extraction schemes for $q_0$ and $\delta$ are required, which we do not include here. To derive them, one employs the strategy that demonstrated following \eqref{LisLidentity}.\\

Considering the statement of Proposition \ref{drifttheorem}, we must have convergence of the series for its truncation to be a valid approximation. For the series to be convergent on $[0,\infty)$, demanding $\delta>\epsilon$ is plausible as the perturbation series grows exponentially otherwise. A sufficient but not necessary criterion to achieve convergence on $[0,\frac1{2\delta}]$ is
\begin{equation}\label{criterion}
    \Gamma:=24e^{\frac{2\epsilon}\omega}|q_0|\left(|z(0)|+\frac1\delta \right)\overset!<1.
\end{equation}
To prove (\ref{criterion}) one applies the variation of parameters formula to (\ref{znODE}) and derives a recursive upper bound $u_n$ for $|z_n|$. Solving the recursion and demanding $\sum_{n\geq 0}u_n\delta^n$ to be convergent then gives (\ref{criterion}).

\section{Simulation}\label{simulations}
We implemented the equations studied above in Mathematica: All differential equations have been numerically solved using the \verb+NDSolve+ function. The following graphics are generated by evaluating the extraction schemes at $t\geq 0$ and plotting the result.  
\subsection{Simple model}
\begin{figure}
    \centering
    \includegraphics[width=0.45\textwidth]{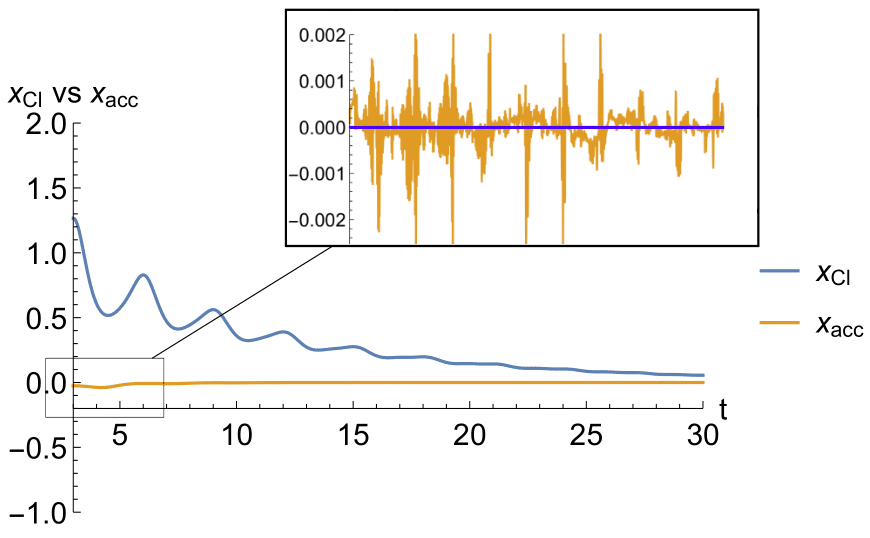}
    \caption{Classical ES vs accelerated ES}
    \label{fig:toyzoom}
\end{figure}

Fig. \ref{fig:toyzoom} shows the classical ES (as depicted in Fig. \ref{noiseloop} without noise) versus the accelerated ES for parameters $T:=3$, $b:=2$, $\epsilon:=.01$ and $L=0$.
The zoomed-in section of the figure shows that the accelerated curve oscillates around $L=0$ with amplitude $\propto\epsilon^2$ as is to be expected from the theory. The initial conditions of the loop are absent in the accelerated scheme for $t\geq 0$. This is due to the extraction scheme using the data points $x(t+kT)$ with $k\leq 3$ (see \eqref{extract1} and \eqref{theaeq}).  \\ 

\begin{figure}[ht]
    \centering
    \includegraphics[width=0.45\textwidth]{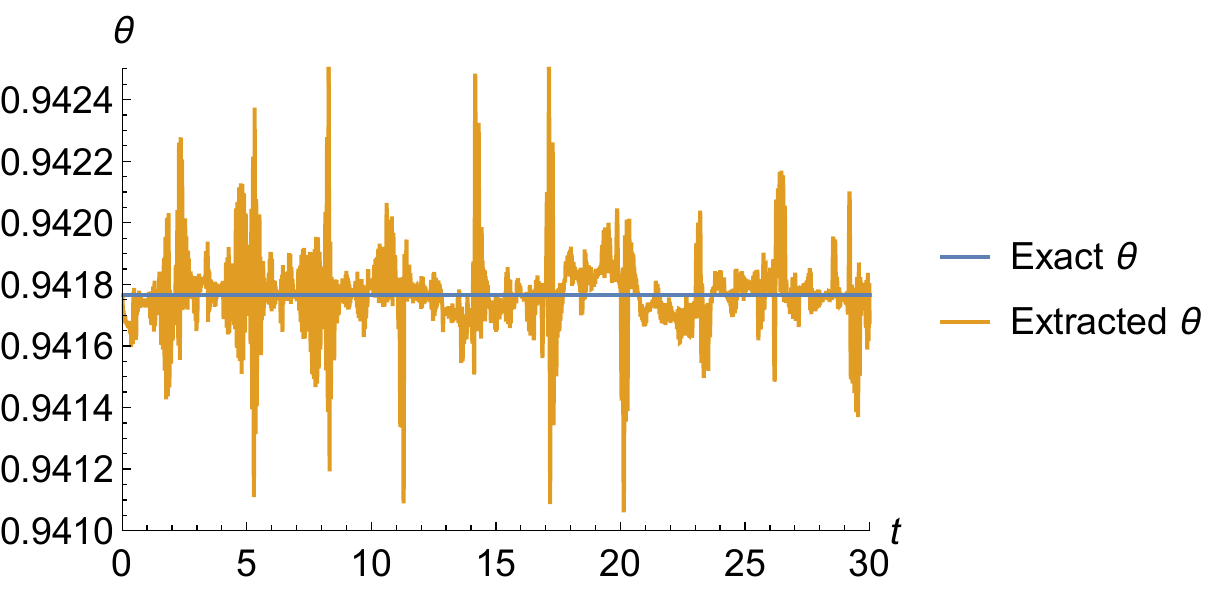}
    \caption{Extraction of $\theta$}
    \label{toythetaextra}
\end{figure}

 Fig. \ref{toythetaextra} demonstrates the extraction of $\theta$ and shows excellent agreement with the exact value $e^{-\epsilon b T}\approx.9418$.\\

\subsection{Including noise}
\begin{figure}
    \centering
    \includegraphics[width=0.45\textwidth]{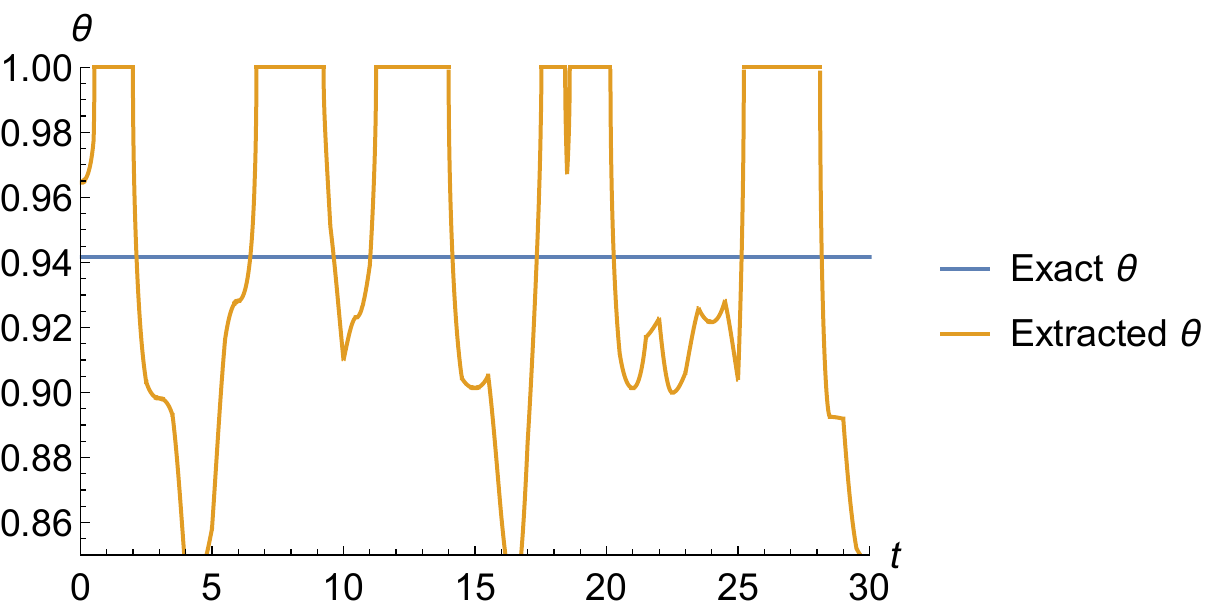}
    \caption{Extraction of $\theta$ ($N_0=\epsilon^2$).}
    \label{NoisethetaO2}
\end{figure}

\begin{figure}
    \centering
    \includegraphics[width=0.45\textwidth]{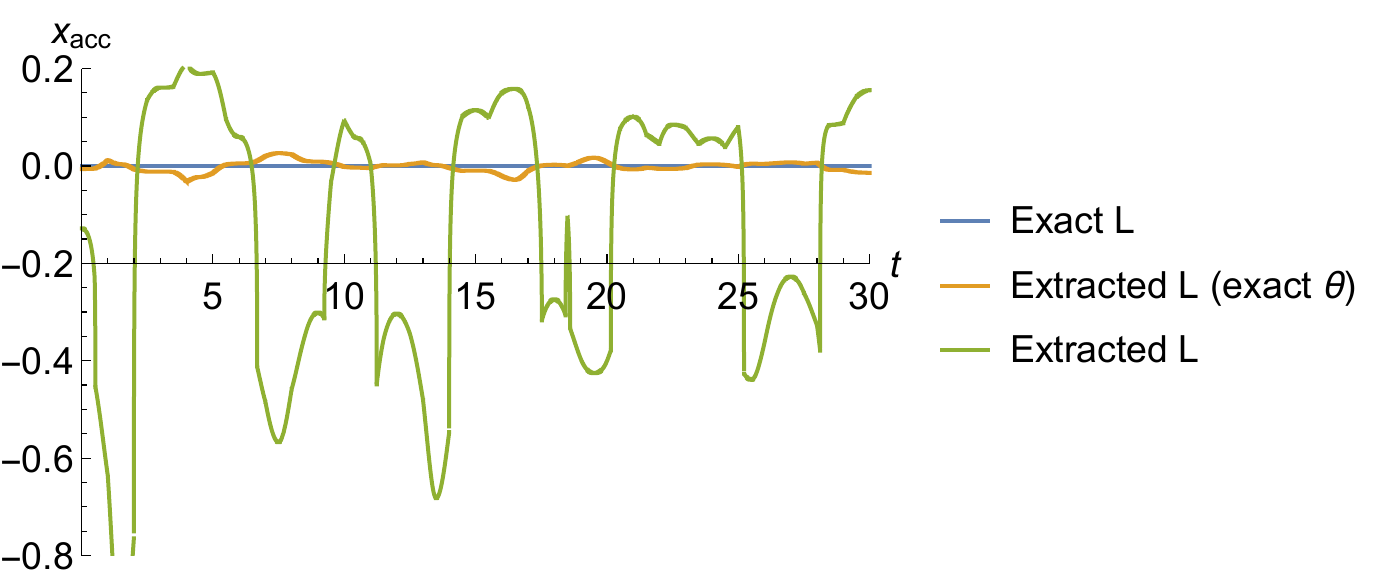}
    \caption{Extraction of $L$ ($N_0=\epsilon^2$).}
    \label{NoiseLO2}
\end{figure}

\begin{figure}
    \centering
    \includegraphics[width=0.45\textwidth]{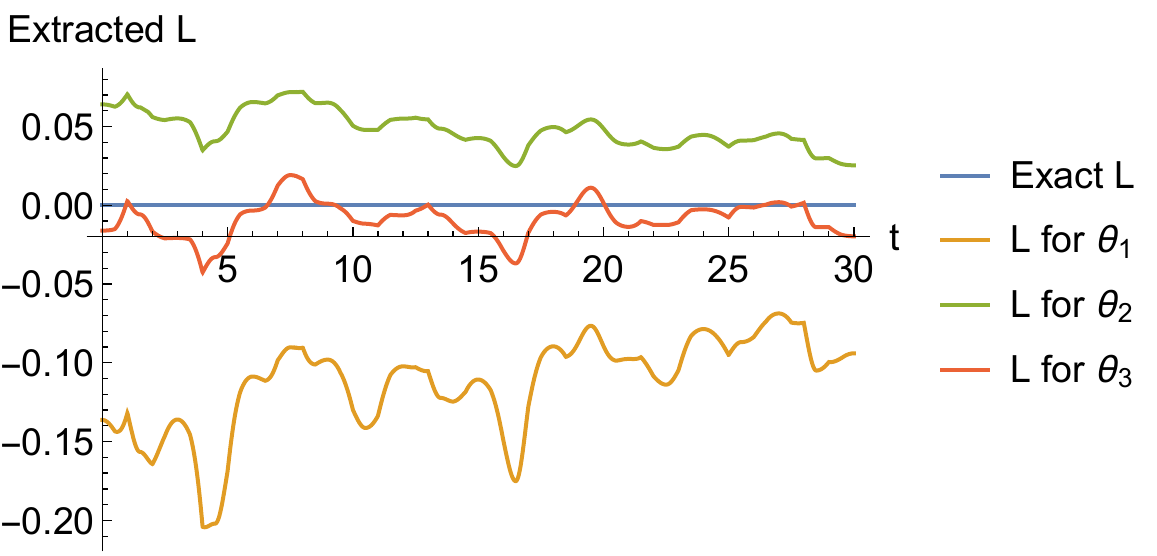}
    \caption{Extraction of $L$ (averaged $\theta$, $N_0=\epsilon^2$).}
    \label{cleanedL}
\end{figure}

We now include the noise block in Fig. \ref{noiseloop}. The noise is realized as a piecewise constant function that takes randomized values in $[-N_0,N_0]$ on intervals of length $dt$. In all following simulations we use  $b=2$, $T=3$, $\epsilon=.01$ and $dt=.5$. 
To explain the following simulation results, we remark that the inclusion of a noise source introduces a new term in (\ref{basicyeq}): 
\begin{align}
&\dot y+\epsilon b(1-\cos(2\omega t))y+by^2\sin(\omega t)\nonumber\\
=&-b\epsilon^2\sin(\omega t)^3-\nu(t)\sin(\omega t)\label{noiseeq}
\end{align}
The analysis in Subsection \ref{basicmodel} is based on dropping terms of order $\epsilon^2$ suggesting that noise of higher amplitude corrupts the method. Indeed, the scheme breaks down for $N_0=\epsilon$. 
Taking $N_0=\epsilon^2$ renders the \emph{noise-term} in (\ref{noiseeq}) to be of order $\epsilon^2$ suggesting the extraction schemes to work. Fig. \ref{NoisethetaO2} and Fig. \ref{NoiseLO2} show the extraction of $\theta$ and $L$ with exact and extracted $\theta$ respectively. The cutoff visible in Fig. \ref{NoisethetaO2} is caused by cutting off $g$ at $g=\frac13$ as larger values lead to complex $\theta$. Extraction of $L$ using the exact value of $\theta$ works fine. However, inclusion of noise causes noticeable oscillations in the extraction of $\theta$ which render the full extraction scheme for $L$ to work poorly. Averaging $\theta$ over time can, however, drastically improve this result. Fig. \ref{cleanedL} shows the extracted value of $L$ that is obtained when using the average value $\theta_k$ of $\theta$ on $[0, kT]$ in \eqref{extract1}.

Smaller $N_0$ such as $N_0=\epsilon^{\frac52}$ render the extraction of $\theta$ accurate enough to extract $L$ without having to resort to averaging procedures. Modifying $dt$ or adding an offset of order at most $\epsilon^2$ to the noise does not change the simulation results.

\subsection{Including a drift}
\begin{figure}[h!]
    \centering
    \includegraphics[width=0.5\textwidth]{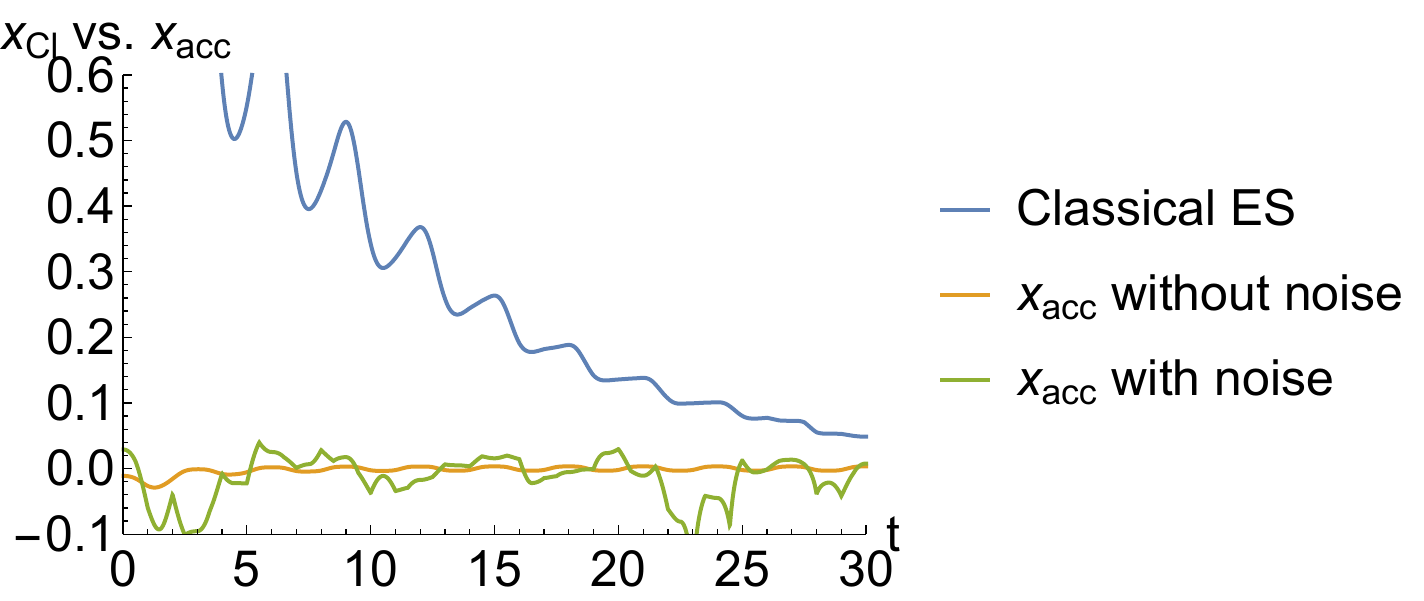}
    \caption{Classical vs accelerated ES.}
    \label{drift1}
\end{figure}

\begin{figure}[h!]
    \centering
    \includegraphics[width=0.5\textwidth]{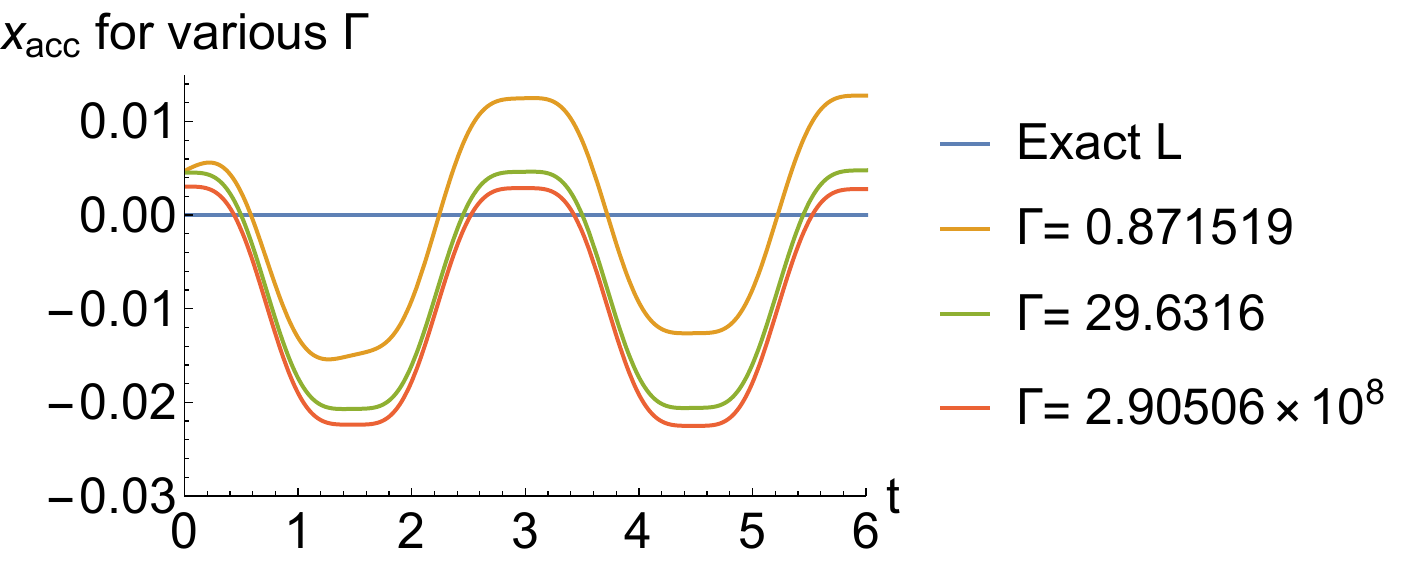}
    \caption{Various values of $\Gamma$/ $\delta$.}
    \label{drift2}
\end{figure}

For all following simulations, we choose $T=3$, $L=0$ and $z(0)=\frac12$. Additionally, taking  $\delta=.4$, $\epsilon=.1$ and $q_0=.01$ gives $\Gamma=.79$ thereby ensuring the scheme to function properly as is verified in Fig. \ref{drift1}. Reusing the terminology from the previous Subsection,  Fig. \ref{drift1} also shows the effect of noise with $N_0=\epsilon^2$ on the scheme. $\Gamma<1$ is, however, not necessary: Taking $\epsilon=.2$, $q_0=.01$ and $\delta\in\set{1,.1,10^{-9}}$ produces accelerated convergence with high values of $\Gamma$ (see Fig. \ref{drift2}).
However, taking $\delta=.1$, $\epsilon=.01$ and e.g. $q_0\in\set{.4,.05}$ shows that that for $\Gamma>1$ the acceleration scheme can in fact break down.

Fig. \ref{drift2} is restricted to $0\leq t\leq 6$ to make the differences between the curves visible. Again, modification of $dt$ and the inclusion of a small offset have no effect on the results.

\section{Summary and outlook}
We have demonstrated how ES loops can be analyzed by considering a perturbation expansion around simpler loops and how the resulting information can be used to derive extraction schemes that speed up the convergence drastically. This statement also holds in comparison to other acceleration schemes, such as fixed-time extremum seeking (see e.g. \cite{poveda2020fixedtime}). The obvious downside of the scheme is that it requires more information about the structure of the system that is to be optimized. The presented scheme is therefore suited to systems of which the physics (but not necessarily the system parameters!) are known and require fast convergence with little oscillations in the steady state, such as in robotics applications. There are still many open questions to be considered: General statements and formal proofs are needed to make the proof of concept presented here more rigorous. This also includes a detailed discussion concerning convergence. Experimental evidence is needed to show the suitability to real-world applications. Finally, additional generalizations such as multidimensional ES are still to be discussed.

\appendix 

\def\thesection{\Alph{section}}
\subsection{Calculus Lemmata}
\label{app:calclem}
\begin{clem}\label{calclem}
Let $L>0$, $1\neq a\in\R^+$ and $y\in C^1(\R)$ such that $y'(x+L)=ay'(x)$. Then 
$$y(x)=\alpha+a^{\frac xL}P(x)$$
for some $\alpha\in \R$ and $L$-periodic $P\in C^1(\R)$.
\end{clem}
\begin{prooof}
We only prove the Lemma for $x\geq 0$. For $x<0$ one argues similarly. Since $(y(x)-ay(x-L))'=0$ there exists some $C\in\R$ such that $y(x)=C+ay(x-L)$. Let $x\geq 0$. There exist unique $n\in\N_0$ and $h\in[0,L)$ such that $x=nL+h$. Using $n=\frac{x-h}L$ we compute
\begin{align*} 
y(x)&=C+ay(x-L)=C(1+a)+a^2y(x-2L)\\
&=...=C(1+a+...+a^{n-1})+a^ny(h)\\
&=C\frac{a^n-1}{a-1}+a^ny(h)\\
&=-\frac{C}{a-1}+a^{\frac xL}a^{-\frac hL}\left(y(h)+\frac C{a-1}\right).
\end{align*}
Setting $\alpha:=-\frac{C}{a-1}$ and $P(x):=a^{-\frac hL}\left(y(h)-\alpha\right)$ we get $y(x)=\alpha+a^{\frac xL}P(x)$. $P$ is $L$-periodic as $h(x+L)=h(x)$ and $P\in C^1$ follows from $P(x)=a^{-\frac xL}(y(x)-\alpha)$.
\end{prooof}

\vspace{-.35cm}
\begin{clem}\label{driftcalclem}
Let $\eta,\omega,a\in\R$, $T:=\frac{2\pi}\omega$, $q\in C^0(\R)$ be $T$-periodic and $y$ be a solution to
$$\dot y(t)+2a\sin^2(\omega t)y(t)=e^{\eta t}q(t).$$
Then $y(t)=e^{-at}p_1(t)+e^{\eta t}p_2(t)$ for some $T-$periodic functions $p_1$ and $p_2$.
\end{clem}
\begin{prooof}
Using $2\sin^2(\omega t)=1-\cos(2\omega t)$ it is readily seen that 
$$\frac d{dt}\left[y(t) e^{at-\frac{a\sin(2\omega t)}{2\omega}}\right]=q(t)e^{\eta t}e^{at-\frac{a\sin(2\omega t)}{2\omega}}.$$
Lemma \ref{calclem} implies the existence of a constant $\rho_0\in\R$ and a $T$-periodic function $\rho(t)$ such that 
$$y(t) e^{at-\frac{a\sin(2\omega t)}{2\omega}}=\rho_0+e^{\eta t}e^{at}\rho(t).$$
This proves the Lemma.
\end{prooof}
\vspace{-.9cm}

\subsection{Proof of Equation (\ref{sol})}
\label{app:eq4}
As described in the paragraphs preceding (\ref{sol}) we study the ODE 
$$\dot y+\epsilon b(1-\cos(2\omega t))y+by^2\sin(\omega t)=0.$$
We put $z:=\frac 1y$ such that $\dot z=-y^{-2}\dot y$ and get
$$\dot z-\epsilon b(1-\cos(2\omega t))z=b\sin(\omega t).$$
Note that $x_0(t)=\exp(-\epsilon b(t-\frac{\sin(2\omega t)}{2\omega}))$ defines an integrating factor for the left hand side. Hence
$$\frac d{dt}\big(z(t)x_0(t)\big )=bx_0(t)\sin(\omega t).$$
Integrating from $0$ to $t$ and abbreviating $z(0)x_0(0)=:C$ yields 
$$z(t)x_0(t)=C+b\int_0^tx_0(s)\sin(\omega s)ds.$$
Equation (\ref{sol}) follows by definition of $z$.

\subsection{Proof of Equation (\ref{theaeq})}\label{thetaeqproof}
Proving that (\ref{theaeq}) is true up to the sign in front of the square root is trivial. To prove that it is `$-$', we use $\theta=e^{-\epsilon bT}\in (0,1)$. We get $(g-1)^2\geq 4g^2$ and thus $-1\leq g\leq\frac13$ as $\theta\in\R$. Additionally, $g=0$ is not possible by \eqref{gdefinition}. Indeed, $\theta\neq 1$ and \eqref{system1} imply $x_0\neq x_1$ and $x_2\neq x_3$. For $g\in[-1,0)$ we have $\frac{1-g}{2g}<0$. Hence
$$\pm\frac1{2g}\sqrt{-4g^2+(g-1)^2}\overset!\geq 0$$
This implies that $-$ is the correct sign. For $g\in(0,\frac13)$ we note that $\frac{1-g}{2g}\geq 1$ and thus
$$\pm\frac1{2g}\sqrt{-4g^2+(g-1)^2}\overset!\leq 0$$
implying again that $-$ is correct. 

\subsection{Proof of Equations (\ref{znODE}) and (\ref{znformula})}
\setcounter{section}{5}
We substitute $z(t)=\sum_{n\geq 0}z_n(t)\delta^n$ into (\ref{zODE}) to get 
\begin{align*}
    \sum_{n=0}^\infty&\left(\dot z_n-2z_n\epsilon\sin^2(\omega t)\right)\delta^n=\sin(\omega t)\\
    &-q(t)\sum_{n=0}^\infty\left[\delta^{n+1}\sum_{j=0}^n z_j z_{n-j}\right].
\end{align*}
Equation (\ref{znODE}) follows by comparing the coefficients of $\delta^n$. Equation (\ref{znformula}) is proven inductively. For $n=0$ it follows by applying Lemma \ref{driftcalclem} to 
$$\dot z_0-2\epsilon\sin^2(\omega t) z_0=\sin(\omega t).$$
Supposing (\ref{znformula}) for $z_0,\hdots,z_n$, it is checked by direct computation that there exist $T$-periodic functions $q_{\kappa,\beta}$ such that
$$q(t)\sum_{j=0}^n z_j z_{n-j}=\sum_{\kappa=1}^{n+1}\left[e^{-\kappa\delta}\sum_{\beta=0}^{\kappa+1}e^{\epsilon\beta t}q_{\kappa,\beta}(t)\right].$$
Using the linearity of (\ref{znODE}) and Lemma \ref{driftcalclem} readily implies (\ref{znformula}) for $z_{n+1}$.

\subsection{Proof of Equation (\ref{criterion})}\label{convergenceproof}
\begin{clem}\label{estimateLem0}
Let $\mu> 0$ and $f\in C^0(\R)$ be positive. Then, for all $t\in[0,\frac1{2\mu}]$
$$\int_0^t e^{-\mu s}f(s)ds\leq 2e^{-\mu t}\int_0^ t f(s)ds.$$
\end{clem}
\begin{proof}
Let $F(t):=\int_0^t e^{-\mu s}f(s)ds$. As $F$ is increasing and $F(0)=0$, we may estimate
\begin{align*}
    \int_0^t e^{-\mu s}f(s)ds&=e^{-\mu t}F(t)+\mu\int_0^ t e^{-\mu s} F(s)ds\\
    &\leq (e^{-\mu t}+\mu t)F(t).
\end{align*}
Using $x\leq e^{-x}$ for $x\leq \frac12$, the Lemma follows.
\end{proof}
\begin{clem}\label{EstimateLem1}
Let $\epsilon>0$, $\omega\in\R$, $R\in C^0(\R)$ and $\xi$ solve $\dot\xi(t)-2\epsilon\sin^2(\omega t)\xi(t)=R(t)$. 
Then, for $t\in[0, \frac1{2\epsilon}]$
$$|\xi(t)|\leq |\xi(0)|e^{\frac\epsilon{2\omega}} e^{\epsilon t}+2e^{\frac\epsilon\omega} \int_0^t |R(s)|ds.$$
\end{clem}
\begin{proof}
It is clear that \vspace{-1ex} 
\begin{align}
\xi(t)=&e^{\epsilon t-\frac\epsilon{2\omega}\sin(2\omega t)}\Big[\xi(0)\nonumber\\
&+\int_0^t e^{-\epsilon s+\frac\epsilon{2\omega}\sin(2\omega s)}R(s)ds\Big]\label{estimate1}.
\end{align}
Estimating the second term using 
 Lemma \ref{estimateLem0} to the second term in \eqref{estimate1} gives the Lemma.
\end{proof}

We now prove \eqref{criterion}.
\begin{proof}
Put $t_0:=\frac1{2\delta}$ and $u_k:=\sup_{0\leq s\leq t_0}|z_k(s)|$. Applying Lemma \ref{EstimateLem1} to \eqref{z0ODE} gives 
\begin{equation}\label{u0estimate}
u_0\leq |z(0)|e^{\frac\epsilon{2\omega}+\epsilon t_0}+2e^{\frac\epsilon{\omega}}t_0=:\alpha_0.
\end{equation}
For $n\geq 0$, applying Lemma \ref{EstimateLem1} to \eqref{znODE} and subsequently using Lemma \ref{estimateLem0} gives
\begin{align*}
    u_{n+1}&\leq 2 e^{\frac\epsilon\omega}|q_0|\sum_{j=0}^{n}\int_0^{t_0} e^{-\delta s} |z_j(s) z_{n-j}(s) |ds\\
    &\leq 4 e^{\frac\epsilon\omega}|q_0|e^{-\delta t_0}\sum_{j=0}^{n}\int_0^{t_0}  |z_j(s) z_{n-j}(s) |ds\\
    &\leq 4 e^{\frac\epsilon\omega}|q_0|e^{-\delta t_0}t_0\sum_{j=0}^{n}u_j u_{n-j}.
\end{align*}
Note $t_0e^{-\delta t_0}\leq\delta^{-1}$, put $C:=4(e\delta)^{-1}e^{\frac\epsilon\omega}|q_0|$ and, for $n\geq 0$, define $\alpha_n$ by\vspace{-2ex}
\begin{equation}\label{alpharecursion}
    \alpha_{n+1}=C\sum_{j=0}^n\alpha_j\alpha_{n-j}.
\end{equation}
An inductive argument shows $u_n\leq\alpha_n$ and hence $\sum_{n\geq 0} z_n\delta ^n$ converges absolutely when $\sum_{n\geq 0}\alpha_n\delta^n$ converges. Consider the generating function $A(x):=\sum_{j\geq 0}\alpha_j x^j$. Using (\ref{alpharecursion}) it is readily checked that $Cx A^2(x)=A(x)-\alpha_0$ and hence 
\begin{equation}\label{Asolution} 
A(x)=\frac{1-\sqrt{1-4C\alpha_0 x}}{2Cx}.
\end{equation}
Expanding (\ref{Asolution}) and using Stirling's approximation gives 
\vspace{-.02cm}
$$\alpha_n\sim\frac{(4C)^n}{\sqrt{\pi}(n+1)^{\frac32}}\alpha_0^{n+1}.$$
\vspace{-.03cm}
Thus, $\sum_{n\geq 0}\alpha_n\delta^n$ converges when $4C\alpha_0\delta<1$. Inserting $\alpha_0$ from (\ref{u0estimate}) gives (\ref{criterion}).
\end{proof}

\bibliographystyle{IEEEtran}
\bibliography{lit.bib}

\begin{thebibliography}{10}
\providecommand{\url}[1]{#1}
\csname url@samestyle\endcsname
\providecommand{\newblock}{\relax}
\providecommand{\bibinfo}[2]{#2}
\providecommand{\BIBentrySTDinterwordspacing}{\spaceskip=0pt\relax}
\providecommand{\BIBentryALTinterwordstretchfactor}{4}
\providecommand{\BIBentryALTinterwordspacing}{\spaceskip=\fontdimen2\font plus
\BIBentryALTinterwordstretchfactor\fontdimen3\font minus
  \fontdimen4\font\relax}
\providecommand{\BIBforeignlanguage}[2]{{%
\expandafter\ifx\csname l@#1\endcsname\relax
\typeout{** WARNING: IEEEtran.bst: No hyphenation pattern has been}%
\typeout{** loaded for the language `#1'. Using the pattern for}%
\typeout{** the default language instead.}%
\else
\language=\csname l@#1\endcsname
\fi
#2}}
\providecommand{\BIBdecl}{\relax}
\BIBdecl

\bibitem{leblanc}
M.~Leblanc, ``Sur i'electrification des chemins de fer au moyen de courants
  alternatifs de frequence elevee,'' \emph{Revue générale de
  l’électricité}, 1922.

\bibitem{book}
K.~B. Ariyur and M.~Krsti{\'{c}}, \emph{Real-Time Optimization by
  Extremum-Seeking Control}.\hskip 1em plus 0.5em minus 0.4em\relax John Wiley
  {\&} Sons, 2003.

\bibitem{Calli2012}
B.~Calli, W.~Caarls, P.~Jonker, and M.~Wisse, ``Comparison of extremum seeking
  control algorithms for robotic applications,'' in \emph{{IEEE}/{RSJ}
  International Conference on Intelligent Robots and Systems}, 2012, pp.
  3195--3202.

\bibitem{Nesic2009}
D.~Ne{\v{s}}i{\'{c}}, ``Extremum seeking control: Convergence analysis,''
  \emph{European Journal of Control}, vol.~15, no. 3-4, pp. 331--347, 2009.

\bibitem{Moura2013}
S.~J. Moura and Y.~A. Chang, ``Lyapunov-based switched extremum seeking for
  photovoltaic power maximization,'' \emph{Control Engineering Practice},
  vol.~21, no.~7, pp. 971--980, 2013.

\bibitem{Malek2016}
H.~Malek and Y.-Q. Chen, ``Fractional order extremum seeking control:
  Performance and stability analysis,'' \emph{{IEEE}/{ASME} Transactions on
  Mechatronics}, vol.~21, no.~3, pp. 1620--1628, 2016.

\bibitem{Haring2016}
M.~A. Haring, ``Extremum-seeking control: convergence improvements and
  asymptotic stability,'' Ph.D. dissertation, NTNU, 2016.

\bibitem{Poveda2020}
J.~I. Poveda and M.~Krsti{\'c}, ``Fixed-time gradient-based extremum seeking,''
  in \emph{American Control Conference}, 2020, pp. 2838--2843.

\bibitem{poveda2020fixedtime}
------, ``Fixed-time newton-like extremum seeking,'' in \emph{{IFAC World
  Congress}}, 2020, pp. 5356--5361.

\bibitem{Richardson}
L.~F. Richardson, ``The approximate arithmetical solution by finite differences
  of physical problems involving differential equations with an application to
  the stresses in a masonry dam,'' \emph{Transactions of the Royal Society of
  London}, vol. Ser. A, no. 210, pp. 307--357, 1910.

\bibitem{example1}
M.~Borinsky, G.~V. Dunne, and M.~Meynig, ``Semiclassical trans-series from the
  perturbative {Hopf}-algebraic {Dyson}-{Schwinger} equations: $\phi^3$ qft in
  6 dimensions,'' \emph{SIGMA}, vol.~17, p. 087 (26 pages), 2021.

\bibitem{example2}
C.~M. Bender, A.~Pelster, and F.~Weissbach, ``Boundary-layer theory,
  strong-coupling series, and large-order behavior,'' \emph{Journal of
  Mathematical Physics}, vol.~43, no.~8, pp. 4202--4220, 2002.

\bibitem{ML}
F.~Bach, ``On the effectiveness of richardson extrapolation in machine
  learning,'' \emph{SIAM Journal on Mathematics and Data Science}, vol.~3,
  no.~4, pp. 1251--1277, 2020.

\bibitem{Brunton2010}
S.~L. Brunton, C.~W. Rowley, S.~R. Kulkarni, and C.~Clarkson, ``Maximum power
  point tracking for photovoltaic optimization using ripple-based extremum
  seeking control,'' \emph{{IEEE} Transactions on Power Electronics}, vol.~25,
  no.~10, pp. 2531--2540, 2010.

\bibitem{bioreactor}
H.-H. Wang, M.~Krsti{\'c}, and G.~Bastin, ``Optimizing bioreactors by extremum
  seeking,'' \emph{International Journal of Adaptive Control and Signal
  Processing}, vol.~13, no.~8, pp. 651--669, 1999.

\bibitem{MLbook}
O.~Calin, \emph{Deep Learning Architectures: A Mathematical Approach}.\hskip
  1em plus 0.5em minus 0.4em\relax Springer, 2020.

\bibitem{benderreview}
C.~M. Bender and C.~Heissenberg, ``Convergent and divergent series in
  physics,'' 2017, {Saalburg Summer School, Germany}.

\bibitem{Amann}
H.~Amann, \emph{Ordinary Differential Equations: An Introduction to Nonlinear
  Analysis}.\hskip 1em plus 0.5em minus 0.4em\relax De Gruyter, 2011.

\bibitem{Bender}
C.~M. Bender and S.~A. Orszag, \emph{Advanced mathematical methods for
  scientists and engineers: I: Asymptotic methods and perturbation
  theory}.\hskip 1em plus 0.5em minus 0.4em\relax Springer, 1999.

\end{thebibliography}

\onecolumn

\begin{center}
    \textbf{\Large Supplementary}
\end{center}

\noindent
\textbf{\large Structure of this Part}\ \\
In the first section, we give more details on the technical Lemmas that are presented in Appendix A. Section 2 provides derivations that have been left out in Section III.A and the corresponding parts of the appendix. Similarly, Section 3 provides derivations that have been left out in Section III.B and the corresponding parts of the appendix. Finally, Section 4 describes how the simulations in Section IV have been generated.
\vspace{1cm}

\addtocounter{section}{1}
\noindent
\setcounter{theorem}{0}
\textbf{\large F. Technical Lemmas}\ \\
\begin{clem}\label{calclem_Supp}
Let $L>0$, $1\neq a\in\R^+$ and $y\in C^1(\R)$ such that $y'(x+L)=ay'(x)$. Then 
$$y(x)=\alpha+a^{\frac xL}P(x)$$
for some $\alpha\in \R$ and $L$-periodic $P\in C^1(\R)$.
\end{clem}
\begin{proof}
We only prove the Lemma for $x\geq 0$. For $x<0$ one argues similarly. Since $(y(x)-ay(x-L))'=0$ there exists some $C\in\R$ such that $y(x)=C+ay(x-L)$. Let $x\geq 0$. There exist unique $n\in\N_0$ and $h\in[0,L)$ such that $x=nL+h$. Using $n=\frac{x-h}L$ we compute
\begin{align} 
y(x)&=C+ay(x-L)\nonumber\\
    &=C(1+a)+a^2y(x-2L)\nonumber\\
&=...\nonumber\\
&=C(1+a+...+a^{n-1})+a^ny(h)\nonumber\\
&=C\frac{a^n-1}{a-1}+a^ny(h)\nonumber\\
&=-\frac{C}{a-1}+a^{\frac xL}a^{-\frac hL}\left(y(h)+\frac C{a-1}\right).\label{lemma1p1g1}
\end{align}
We now define 
$$\alpha:=-\frac{C}{a-1}\hspace{.5cm}\textrm{and}\hspace{.5cm}P(x):=a^{-\frac hL}\left(y(h)-\alpha\right).$$
Inserting these definitions into Equation \eqref{lemma1p1g1}, we get
\begin{equation}\label{lemma1p1g2}
y(x)=\alpha+a^{\frac xL}P(x)
\end{equation}
$P$ is $L$-periodic. Indeed, if $x=nL+h$, then $x+L=(n+1)x+h$ and hence 
$$P(x+L)=a^{-\frac hL}(y(h)-\alpha)=P(x).$$
To prove $P\in C^1$ we rewrite Equation \eqref{lemma1p1g2} as 
$$P(x)=a^{-\frac xL}(y(x)-\alpha)\in C^1.$$
In the last step we have used the regularity of $y$. 
\end{proof}

\begin{clem}\label{driftcalclem_Supp}
Let $\eta,\omega,a\in\R$, $T:=\frac{2\pi}\omega$, $q\in C^0(\R)$ be $T$-periodic and $y$ be a solution to
$$\dot y(t)+2a\sin^2(\omega t)=e^{\eta t}q(t).$$
Then $y(t)=e^{-at}p_1(t)+e^{\eta t}p_2(t)$ for some $T-$periodic functions $p_1$ and $p_2$. Additionally, $p_2=0$ if $q=0$. 
\end{clem}
\begin{proof}
Using $2\sin^2(\omega t)=1-\cos(2\omega t)$, we compute
\begin{align}
    \frac d{dt}\left[y(t) e^{at-\frac{a\sin(2\omega t)}{2\omega}}\right]&=e^{at-\frac{a\sin(2\omega t)}{2\omega}}\left[\dot y(t) +y(t)\frac{d}{dt}\left(at-\frac{a\sin(2\omega t)}{2\omega}\right)\right]\nonumber\\
    &=e^{at-\frac{a\sin(2\omega t)}{2\omega}}\left[
    \dot y(t) +y(t)\left(a-a\cos(2\omega t)\right)
    \right]\nonumber\\
    &=e^{at-\frac{a\sin(2\omega t)}{2\omega}}\left[
    \dot y(t)+ 2a\sin^2(\omega t)y
    \right]\nonumber\\
    &=e^{at-\frac{a\sin(2\omega t)}{2\omega}}e^{\eta t} q(t).\label{supp1}
\end{align}
In the last step we have used the ODE that $y$ solves.
If $q=0$, the right hand side in Equation \eqref{supp1} vanishes and we deduce that there is a constant $C$ such that 
$$y(t)e^{at-\frac{a\sin(2\omega t)}{2\omega}}=C.$$
Hence 
$$y(t)=e^{- at} C e^{\frac{a\sin(2\omega t)}{2\omega}}$$
which is the claimed formula. For general $q$ we define
\begin{equation}\label{supp2}
f(t):=e^{at-\frac{a\sin(2\omega t)}{2\omega}}e^{\eta t} q(t)\hspace{.5cm}\textrm{and}\hspace{.5cm}Y(t):= \frac d{dt}\left[y(t) e^{at-\frac{a\sin(2\omega t)}{2\omega}}\right].
\end{equation}
Note that 
$$f(t+T)=e^{\eta(t+T)}e^{a(t+T)}e^{-a\frac{\sin(2\omega (t+T))}{2\omega}}=e^{(a+\eta)T}f(t).$$ 
Using Equations \eqref{supp1} and \eqref{supp2} we get 
$$ Y'(t+T)-e^{(a+\eta)T}Y'(t)=f(t+T)-e^{(a+\eta)T}f(t)=0.$$
Lemma \ref{calclem_Supp} implies the existence of a constant $\rho_0\in\R$ and a $T$-periodic function $\rho(t)$ such that 
$$y(t) e^{at-\frac{a\sin(2\omega t)}{2\omega}}=Y(t)=\rho_0+e^{\eta t}e^{at}\rho(t).$$
This yields the claimed formula:
$$y(t)=e^{-at}\rho_0e^{\frac{a\sin(2\omega t)}{2\omega}}+e^{\eta t}\rho(t)e^{\frac{a\sin(2\omega t)}{2\omega}}$$
\end{proof}

\noindent
\textbf{Proof of Equation (8)}\ \\
We define 
\begin{equation}\label{suppx0def}
x_0(t)\exp\left[-\epsilon b\left(t-\frac{\sin(2\omega t)}{2\omega}\right)\right]
\end{equation}
\begin{clem}\label{supplemma1}
Let $\epsilon,b,\omega\in\R$ and $y$ be a solution of 
$$\dot y+\epsilon b(1-\cos(2\omega t))y+by^2\sin(\omega t)=0.$$
Then there exists a constant $C$ such that 
$$y(t)=\frac{x_0(t)}{C+b\int_0^t \sin(\omega s)x_0(s)ds}.$$
\end{clem}
\begin{proof}
We put $z:=\frac 1y$ such that $\dot z=-y^{-2}\dot y$ or equivalently $\dot y=-z^{-2}\dot z$. This gives
\begin{align*}
    \dot z-\epsilon b(1-\cos(2\omega t))z&=-y^{-2}\dot y-\epsilon b(1-\cos(2\omega t))y^{-1}\\
    &=-y^{-2}\left(\dot y+\epsilon b(1-\cos(2\omega t))y\right)\\
    &=-y^{-2}\left(-by^2\sin(\omega t)\right)\\
   &= b\sin(\omega t).
\end{align*}
Using this ODE for $z$ we compute
\begin{align*}
    \frac d{dt}\big(z(t)x_0(t)\big )&=\dot z(t)x_0(t)+z(t)\dot x_0(t)\\
        &=x_0(t)\left[\dot z(t)+z(t)\left(-\epsilon b +\epsilon b \cos(2\omega t)\right)\right]\\
        &=x_0(t)\left[\dot z(t)-\epsilon b z(t)\left(1- \cos(2\omega t)\right)\right]\\
        &=x_0(t)b\sin(\omega t).
\end{align*}
Integrating from $0$ to $t$ and abbreviating $z(0)x_0(0)=:C$ yields 
$$z(t)x_0(t)=C+b\int_0^tx_0(s)\sin(\omega s)ds.$$
The Lemma follows by inserting $z(t)=y(t)^{-1}$. 
\end{proof}

\newpage
\addtocounter{section}{1}
\noindent
\textbf{\large G. Supplementary Details to Section III.A}\\
\textbf{Proof of Equation (10)}\ \\
\noindent
We have 
$$x_n:=x(t+nT)=L+\frac{\theta^n x_0(t)}{\tilde C+\theta^n X(t)}.$$
Let $a:=x_0(t)$, $b:=X(t)$, $c:=\tilde C$ and $y_n:=x_n-L$. We compute
\begin{align*}
    (y_0-y_1)y_2&=\left(\frac{a}{b+c}-\frac{a\theta}{c+b\theta}\right)\frac{a\theta^2}{c+b\theta^2}\\
     &=\left(\frac{ac+ab\theta-a\theta b-a\theta c}{(b+c)(c+b\theta)}\right)\frac{a\theta^2}{c+b\theta^2}\\
     &=\frac{ac-a\theta c}{(b+c)(c+b\theta)}\frac{a\theta^2}{c+b\theta^2}.
\end{align*}
Next, we compute 
\begin{align*}
    \theta(y_2-y_1)y_0&=\theta\left(\frac{a\theta^2}{b\theta^2+c}-\frac{a\theta}{c+b\theta}\right)\frac{a}{c+b}\\
    &=\left(\frac{a\theta^2 c+a\theta^3b-ab\theta^3-ac\theta}{(c+b\theta^2)(c+b\theta)}
    \right)\frac{\theta a}{c+b}\\
    &=\frac{a\theta^2 c-ac\theta}{(c+b\theta^2)(c+b\theta)}
    \frac{\theta a}{c+b}\\
    &=\frac{a\theta c-ac}{(c+b\theta^2)(c+b\theta)}
    \frac{\theta^2 a}{c+b}.
\end{align*}
Combining both equations we get 
$$0=(y_0-y_1)y_2+\theta (y_2-y_1)y_0.$$
Since $y_n=x_n-L$ we have $y_k-y_l=x_k-x_l$. Consequently 
\begin{align*}
    0&=(y_0-y_1)y_2+\theta (y_2-y_1)y_0\\
    &=(x_0-x_1)y_2+\theta (x_2-x_1)y_0\\
    &=(x_0-x_1)x_2+\theta (x_2-x_1)x_0-L\left((x_0-x_1)+\theta(x_2-x_1)\right).
\end{align*}
Rearranging gives 
$$L=\frac{(x_0-x_1)x_2+\theta (x_2-x_1)x_0}{x_0-(1+\theta) x_1+\theta x_2}.$$

\noindent
\textbf{Proof of Equation (13)}\ \\
We consider the Equation 
\begin{equation}\label{supp3}
\frac{(x_0-x_1)x_2+\theta x_0(x_2-x_1)}{x_0-(1+\theta) x_1+\theta x_2}
=
\frac{(x_1-x_2)x_3+\theta x_1(x_3-x_2)}{x_1-(1+\theta) x_2+\theta x_3}
\end{equation}
where we recall that $\theta=e^{-\epsilon b T}\in(0,1)$, $x_n=x(t+nT)$ and
\begin{equation}\label{differencevorbereitung}
x_n=x(t+nT)=L+\frac{x_0(t+nT)}{\tilde C+X(t+nT)}=L+\frac{\theta^n x_0(t)}{\tilde C+\theta^n X(t)}.
\end{equation}
$X(t)$ is a $T$-periodic function and $x_0(t)$ is as in Equation \eqref{suppx0def}.
Note that $x_n\rightarrow L$ as $n\rightarrow \infty$. This implies that $\tilde C\neq 0$ as otherwise 
$$x_n=L+\frac{\theta^n x_0(t)}{\theta^n X(t)}=L+\frac{x_0(t)}{X(t)}\not\rightarrow L$$
as $x_0(t)>0$.
Using Equation \eqref{differencevorbereitung}, we get for $n\geq 0$ and $k\geq 1$
\begin{align}
    x_{n+k}-x_n&=\frac{\theta^{n+k} x_0(t)}{\tilde C+\theta^{n+k} X(t)}-\frac{\theta^n x_0(t)}{\tilde C+\theta^n X(t)}\nonumber\\
    &=\theta^k\frac{\theta^{n} x_0(t)}{\tilde C+\theta^n X(t)}\frac{\tilde C+\theta^n X(t)}{\tilde C+\theta^{n+k} X(t)}-\frac{\theta^n x_0(t)}{\tilde C+\theta^n X(t)}\nonumber\\
     &=\frac{\theta^{n} x_0(t)}{\tilde C+\theta^n X(t)}\left[\theta^k\frac{\tilde C+\theta^n X(t)}{\tilde C+\theta^{n+k} X(t)}-1\right]\nonumber\\
     &=\frac{\theta^{n} x_0(t)}{\tilde C+\theta^n X(t)}
     \left[\frac{\theta^k\tilde C+\theta^{n+k} X(t)}{\tilde C+\theta^{n+k} X(t)}-\frac{\tilde C+\theta^{n+k} X(t)}{\tilde C+\theta^{n+k} X(t)}\right]\nonumber\\
     &=\frac{\theta^{n} x_0(t)}{\tilde C+\theta^n X(t)}
    \frac{\tilde C(\theta^k-1)}{\tilde C+\theta^{n+k} X(t)}\nonumber\\
    &\neq 0.\label{suppdiffneq0}
\end{align} 
We prove:
\begin{equation}\label{supp4}
    \theta=\frac{1-g}{2g}-\frac1{2g}\sqrt{-4g^2+(g-1)^2}
\end{equation}
\begin{proof}
We rewrite Equation Equation \eqref{supp3}:
$$
\bigg((x_0-x_1)x_2+\theta x_0(x_2-x_1)\bigg)\bigg({x_1-(1+\theta) x_2+\theta x_3}\bigg)
=
\bigg((x_1-x_2)x_3+\theta x_1(x_3-x_2)\bigg)\bigg(x_0-(1+\theta) x_1+\theta x_2\bigg)
$$
We further simplify by collecting the terms with and without $\theta$'s in the large parenthesis:
$$
\bigg((x_0-x_1)x_2+\theta x_0(x_2-x_1)\bigg)\bigg(x_1-x_2+\theta (x_3-x_2)\bigg)
=
\bigg((x_1-x_2)x_3+\theta x_1(x_3-x_2)\bigg)\bigg(x_0-x_1\theta (x_2-x_1)\bigg)
$$
Next we expand:
\begin{align*}
    &(x_0-x_1)x_2(x_1-x_2)+\theta(x_0(x_2-x_1)(x_1-x_2)+(x_0-x_1)x_2(x_3-x_2))+\theta^2x_0(x_2-x_1) (x_3-x_2)\\
    =&  (x_1-x_2)x_3(x_0-x_1)+\theta(x_1(x_3-x_2)(x_0-x_1)+(x_1-x_2)x_3(x_2-x_1))+\theta^2x_1(x_3-x_2)(x_2-x_1)
\end{align*}
We subtract all terms in the second line and get:
\begin{align*}
    0=&(x_0-x_1)(x_1-x_2)(x_2-x_3)\\
        &+\theta\left[-(x_0-x_3)(x_1-x_2)^2+(x_0-x_1)(x_3-x_2)(x_2-x_1)\right]\\
        &+\theta^2(x_0-x_1)(x_1-x_2)(x_2-x_3)
\end{align*}
Considering Equation \eqref{suppdiffneq0}, we may divide by $(x_1-x_2)^2(x_0-x_3)$ and get 
$$\frac{(x_0-x_1)(x_2-x_3)}{(x_1-x_2)(x_0-x_3)}(1+\theta+\theta^2)-\theta=0.$$
We define 
$$g:=\frac{(x_0-x_1)(x_2-x_3)}{(x_1-x_2)(x_0-x_3)}.$$
so that $g\theta^2+(g-1)\theta+g=0$. Hence 
$$\theta=\frac{1-g}2\pm\sqrt{\left(\frac{1-g}{2g}\right)^2-1}
\overset!=\frac{1-g}2\pm\frac1{2g}\sqrt{(1-g)^2-4g^2}.$$
In the last step (marked by $!$) we have pulled out $(2g)^{-2}$ out of the square root and written the factor $(2g)^{-1}$ in front of it. Really, we have to write $|2g|^{-1}$. However, we can absorb the potential sign  difference in the still ambiguous $\pm$. Only now we determine the correct sign. To prove that is is `$-$', we use $\theta=e^{-\epsilon bT}\in (0,1)$. In particular, $\theta\in\R$ and so $(g-1)^2\geq 4g^2$, which implies $-1\leq g\leq\frac13$ . Additionally, due to Equation \eqref{suppdiffneq0}, we deduce $g\neq 0$. Now we distinguish two cases.
\newpage
\begin{enumerate}
    \item For $g\in[-1,0)$ we have $\frac{1-g}{2g}<0$. As $\theta\in(0,1)$ we deduce
    $$\pm\frac1{2g}\sqrt{-4g^2+(g-1)^2}\overset!\geq 0.$$
    This implies that $-$ is the correct sign.
    \item  For $g\in(0,\frac13)$ we note that $\frac{1-g}{2g}\geq 1$.  As $\theta\in(0,1)$ we deduce
$$\pm\frac1{2g}\sqrt{-4g^2+(g-1)^2}\overset!\leq 0.$$
Again, this implies that $-$ is the correct sign.
\end{enumerate} 
\end{proof}

\addtocounter{section}{1}
\noindent
\textbf{\large H. Supplementary Details to Section III.B}\ \\
For parameters $b,q_0,\omega\in\R$, $\delta,\epsilon>0$ and we consider the Equation 
\begin{equation}\label{suppzODE}
\dot z-2\epsilon \sin^2(\omega t) z+\delta q(t) z^2 =\sin(\omega t).
\end{equation}
where $q(t)=q_0e^{-\delta t}$. We treat $\delta$ as a perturbative parameter and propose the ansatz 
\begin{equation}\label{supppertuseries}
z(t)=\sum_{n=0}^\infty z_n(t)\delta^n
\end{equation}
with initial values $z_0(0)=z(0)$ and $z_n(0)=0$ for all $n\geq 1$.\\
\ \\
\noindent
\textbf{Proof of Equations (17) and (18)}\ \\

 We claim that the following Equations follow:
\begin{equation}\label{z0ODE_Supp}
\left\{
    \begin{aligned}
    &\dot z_0-2\epsilon\sin^2(\omega t)z_0=\sin(\omega t),\\
    &z_0(0)=z(0).
    \end{aligned}
    \right.
\end{equation}
For $n\geq 1$:
\begin{equation}\label{znODE_Supp}
\left\{
    \begin{aligned}
    &\dot z_n-2\epsilon\sin^2(\omega t)z_n=-q(t)\sum_{j=0}^{n-1}z_jz_{n-1-j},\\
    &z_n(0)=0.
    \end{aligned}
    \right.
\end{equation}
\begin{proof}
Inserting the ansatz into Equation \eqref{suppzODE} gives 
$$\sum_{n=0}^\infty \left(\dot z_n-2z_n \epsilon\sin^2(\omega t)\right) \delta^n +\delta q(t)\left(\sum_{l=0}^\infty z_l\delta^l\right)\left(\sum_{k=0}^\infty z_k\delta ^k\right)=\sin(\omega t).$$
We write 
$$\delta q(t)\left(\sum_{l=0}^\infty z_l\delta^l\right)\left(\sum_{k=0}^\infty z_k\delta ^k\right)
= 
q(t)\delta \sum_{n=0}^\infty \left(\delta^n\sum_{a=0}^n z_a z_{n-a}\right)
=
q(t) \sum_{n=0}^\infty \left(\delta^{n+1}\sum_{a=0}^n z_a z_{n-a}\right).$$
Inserting gives 
$$\sum_{n=0}^\infty \left(\dot z_n-2z_n \epsilon\sin^2(\omega t)\right) \delta^n =\sin(\omega t)-q(t) \sum_{n=0}^\infty \left(\delta^{n+1}\sum_{a=0}^n z_a z_{n-a}\right).$$
Comparing coefficients gives Equations \eqref{z0ODE_Supp} and \eqref{znODE_Supp}.
\end{proof}

\setcounter{theorem}{0}
\begin{clem}\label{supplemma2}
There exist $T$-periodic functions $p_0^{(0)}$ and $p_1^{(0)}$ such that 
$$z_0(t)=p_0^{(0)}(t) +e^{\epsilon t} p_1^{(0)}(t).$$
Further, for $n\geq 1$, $1\leq j\leq n$ and $0\leq k\leq j+1$ there exist $T$-periodic functions $p^{(n)}_0$ and $p_{jk}^{(n)}$ such that 
$$z_n(t)=e^{\epsilon t}p^{(n)}_0(t)+\sum_{j=1}^n \sum_{k=0}^{j+1}e^{(k\epsilon-j\delta) t}p^{(n)}_{jk}(t).$$
\end{clem}

\begin{proof}
For $n=0$ we can apply Lemma \ref{driftcalclem_Supp} with $y\rightarrow z_0$, $a\rightarrow -\epsilon$, $\eta\rightarrow 0$ and $q(t)\rightarrow \sin(\omega t)$ to obtain the claimed formula. 
For $n\geq 1$ we argue by induction. First we consider $n=1$. We have 
$$\dot z_1-2\epsilon \sin^2(\omega t) z_1=-q(t)z_0^2.$$
We insert $z_0(t)=p_0^{(0)}(t) +e^{\epsilon t} p_1^{(0)}(t)$ and use $q(t)=q_0 e^{-\delta t}$ to get 
$$\dot z_1-2\epsilon \sin^2(\omega t) z_1=-q_0 e^{-\delta t}\left[\left(p_0^{(0)}(t) \right)^2+e^{2\epsilon t}\left( p_1^{(0)}(t)\right)^2+2e^{\epsilon t}p_0^{(0)}(t) p_1^{(0)}(t)\right].$$
The general solution to this Equation is given by 
\begin{equation}\label{z1decomposition}
z_1(t)=z_{1,h}(t)+\sum_{k=0}^2 z_{1,k}(t).
\end{equation}
Here $z_{1,h}$ denotes a homogeneous solution and for $k=0,1,2$ the function $z_{1,k}$ is any solution of 
$$\dot z_{1,k}-2\epsilon \sin^2(\omega t) z_{1,k}=e^{-\delta t}e^{k\epsilon t} P_{1,k}(t)$$
where 
$$P_{1,0}:=-q_0\left(p_0^{(0)}\right)^2,\hspace{.5cm} P_{1,1}=-2q_0 p_0^{(0)}p_1^{(0)}\hspace{.5cm}\textrm{and}\hspace{.5cm} P_{1,2}:=-q_0\left(p_1^{(0)}\right)^2.$$
We can apply Lemma \ref{driftcalclem_Supp} to obtain $T$-periodic functions $\pi_*$ ($*$ denotes arbitrary indices) such that
\begin{align*}
    &z_h(t)=e^{\epsilon t} \pi_h(t),\\
    &z_k(t)=e^{\epsilon t}\pi_{k,1}(t)+e^{k\epsilon t-\delta t}\pi_{k,2}(t).
\end{align*}
Using Equation \eqref{z1decomposition}, we get the claimed formula for $z_1$. \\

Now we consider the inductive step $n\rightarrow n+1$. Assume that the formulas for $z_k$ with $0\leq k\leq n$ are already proven. We have to compute  
$$\sum_{a=0}^n z_a z_{n-a}=2 z_0 z_n+\sum_{a=1}^{n-1} z_a z_{n-a}.$$
For $1\leq a\leq n-1$ we have the formulas 
\begin{align*}
    z_a(t)=e^{\epsilon t}p^{(a)}_0(t)+\sum_{j=1}^a \sum_{k=0}^{j+1}e^{(k\epsilon-j\delta) t}p^{(a)}_{jk}(t)\\
    z_{n-a}(t)=e^{\epsilon t}p^{(n-a)}_0(t)+\sum_{j=1}^{n-a} \sum_{k=0}^{j+1}e^{(k\epsilon-j\delta) t}p^{(n-a)}_{jk}(t).
\end{align*}
Multiplying gives 
\begin{align*}
    z_az_{n-a}&=e^{2\epsilon t}p^{(a)}_0p^{(n-a)}_0\\
            &+e^{\epsilon t}p^{(a)}_0(t)\sum_{j=1}^{n-a} \sum_{k=0}^{j+1}e^{(k\epsilon-j\delta) t}p^{(n-a)}_{jk}\\
            &+e^{\epsilon t}p^{(n-a)}_0\sum_{j=1}^a \sum_{k=0}^{j+1}e^{(k\epsilon-j\delta) t}p^{(a)}_{jk}\\
            &+\left(\sum_{j=1}^a \sum_{k=0}^{j+1}e^{(k\epsilon-j\delta) t}p^{(a)}_{jk}\right)\left(\sum_{j=1}^{n-a} \sum_{k=0}^{j+1}e^{(k\epsilon-j\delta) t}p^{(n-a)}_{jk}\right)
\end{align*}
This sum is a linear combination of $e^{-\kappa\delta t}$ with $\kappa=0,...,n$. The coefficients  of $e^{-\kappa\delta t}$ are linear combinations of functions of the form $e^{j\epsilon t}p(t)$ where $p$ stands for a general $T$-periodic function and $0\leq j\leq \kappa+2$. Consequently
$$\sum_{a=1}^{n-1} z_a z_{n-a}=\sum_{\kappa=0}^n\left[ e^{-\kappa\delta t}\sum_{j=0}^{\kappa+2}\pi_{\kappa j}(t) e^{j\epsilon t}\right]$$
where $\pi_{\kappa j}$ are some $T$-periodic functions. 
Now, we compute 
\begin{align*}
    z_0 z_n=&(p_0^{(0)}+e^{\epsilon t}p_1^{(0)})\left(e^{\epsilon t}p^{(n)}_0(t)+\sum_{j=1}^n \sum_{k=0}^{j+1}e^{(k\epsilon-j\delta) t}p^{(n)}_{jk}(t)\right)\\
    =&e^{\epsilon t}p_0^{(0)}p^{(n)}_0(t)+\sum_{j=1}^n \sum_{k=0}^{j+1}e^{(k\epsilon-j\delta) t}p_0^{(0)}p^{(n)}_{jk}(t)\\
    &+e^{2\epsilon t}p_1^{(0)}p^{(n)}_0(t)+\sum_{j=1}^n \sum_{k=0}^{j+1}e^{((k+1)\epsilon-j\delta) t}p_1^{(0)}p^{(n)}_{jk}(t).
\end{align*}
Again, this is a linear combination of $e^{-\kappa\delta t}$ where $0\leq \kappa\leq n$. The coefficients of $e^{-\kappa\delta t}$ are again linear combinations of functions $e^{j\epsilon t}p(t)$ where $0\leq j\leq \kappa+2$ and $p$ stands for a general $T$-periodic function. Therefore we have shown that 
$$e^{-\delta t}\sum_{a=0}^n z_a z_{n-a}
=\sum_{\kappa=0}^ {n}e^{-(\kappa+1)\delta t}\sum_{j=0}^{\kappa+2}\tilde\pi_{\kappa j}(t) e^{j\epsilon t}
=\sum_{\kappa=1}^ {n+1}\left[e^{-\kappa\delta t}\sum_{j=0}^{\kappa+1}\pi_{\kappa j}(t) e^{j\epsilon t}\right]
$$
for some, potentially new. $T$-periodic functions $ \pi_{\kappa j}$. Absorbing $-q_0$ into the definition of the functions $\pi_{\kappa j}$ we get 

\begin{equation}\label{inductivestepode}
\dot z_{n+1}+2\epsilon\sin^2(\omega t)=-q(t)\sum_{a=0}^n z_az_{n-a}=\sum_{\kappa=1}^ {n+1}\left[e^{-\kappa\delta t}\sum_{j=0}^{\kappa+1}\pi_{\kappa j}(t) e^{j\epsilon t}\right].
\end{equation}
We now argue as we did for $n=1$ and write 
$$z_{n+1}=z_{n+1,h}+\sum_{\kappa=1}^{n+1}\sum_{j=0}^{\kappa+1}z_{\kappa,j}(t)$$
where $z_{n+1,h}$ is a homogeneous solution to Equation \eqref{inductivestepode} and for $1\leq\kappa\leq n+1$ and $0\leq j\leq \kappa+1$ 
$$\dot z_{\kappa,j}+2\epsilon\sin^2(\omega t)z_{\kappa,j}=e^{-\kappa\delta t}e^{j\epsilon t}\pi_{\kappa j}.$$
Using Lemma \ref{driftcalclem_Supp} and resumming we deduce that there exist periodic functions $p^{(n+1)}_0$ and $p^{(n+1)}_{jk}$ such that
$$z_{n+1}(t)=e^{\epsilon t}p^{(n+1)}_0(t)+\sum_{j=1}^{n+1} \sum_{k=0}^{j+1}e^{(k\epsilon-j\delta) t}p^{(n+1)}_{jk}(t).$$
This finishes the inductive argument. 
\end{proof}

The last step in the proof of Proposition III.2 is to resum the perturbative series. We compute 
\begin{align*}
    z(t)=&\sum_{n=0}^\infty z_n(t)\delta^n\\
    =&p_0^{(0)}+e^{\epsilon t}p_1^{(0)}+\sum_{n=1}^\infty \delta^n \left[
        e^{\epsilon t}p^{(n)}_0(t)+\sum_{j=1}^n \sum_{k=0}^{j+1}e^{(k\epsilon-j\delta) t}p^{(n)}_{jk}(t)
    \right]\\
=&p_0^{(0)}+e^{\epsilon t}\left[p_1^{(0)}+\delta\sum_{n=1}^\infty p_0^{(n)}\right]
+\sum_{n=1}^\infty \sum_{j=1}^n\sum_{k=0}^{j+1}\delta^n e^{(k\epsilon-j\delta) t}p^{(n+1)}_{jk}(t)\\
=&p_0^{(0)}+e^{\epsilon t}\left[p_1^{(0)}+\delta\sum_{n=1}^\infty p_0^{(n)}\right]
+\sum_{j=1}^\infty \sum_{n=j}^\infty\sum_{k=0}^{j+1}\delta^n e^{(k\epsilon-j\delta) t}p^{(n+1)}_{jk}(t)\\
=&p_0^{(0)}+e^{\epsilon t}\left[p_1^{(0)}+\delta\sum_{n=1}^\infty p_0^{(n)}\right]
+\sum_{j=1}^\infty\left[ \delta^j e^{-j\delta t} \sum_{k=0}^{j+1}\left(e^{k\epsilon t}\sum_{n=j}^\infty
\delta^{n-j}p^{(n+1)}_{jk}(t)\right)\right].
\end{align*}
For $j\geq 1$ and $0\leq k\leq j+1$ we define
$$p_{jk}(t):=\sum_{n=j}^\infty
\delta^{n-j}p^{(n+1)}_{jk}(t).$$
Note that $n-j\geq 0$, so $p_{jk}=\mathcal O(1)$ with respect to $\delta$. In particular, $p_{jk}$ do not blow up when $\delta\rightarrow 0$. Additionally, we put 
$$p_{00}:=p_0^{(0)}\hspace{.5cm}\textrm{and}\hspace{.5cm}p_{01}:=p_1^{(0)}+\sum_{n=1}^\infty p_0^{(n)}.$$
Clearly, for all $0\leq j$ and $0\leq k\leq j+1$, the functions $p_{jk}$ are $T$-periodic. Also, we get
\begin{align*}
    z(t)=&p_{00}+e^{\epsilon t}p_{01}
+\sum_{j=1}^\infty\left[ \delta^j e^{-j\delta t} \sum_{k=0}^{j+1}e^{k\epsilon t}p_{jk}(t)\right]\\
=&\sum_{j=0}^\infty\left[ \delta^j e^{-j\delta t} \sum_{k=0}^{j+1}e^{k\epsilon t}p_{jk}(t)\right].
\end{align*}

\noindent
\textbf{Proof of Equation (20)}\ \\
\begin{clem}\label{estimateLem0_Supp}
Let $\mu> 0$ and $f\in C^0(\R)$ be positive. Then, for all $t\in[0,\frac1{2\mu}]$
$$\int_0^t e^{-\mu s}f(s)ds\leq 2e^{-\mu t}\int_0^ t f(s)ds.$$
\end{clem}
\begin{proof}
Let $F(t):=\int_0^t e^{-\mu s}f(s)ds$. As $F$ is increasing and $F(0)=0$, we may estimate
\begin{align*}
    \int_0^t e^{-\mu s}f(s)ds&=\int_0^t e^{-\mu s} F'(s)ds\\
    &=e^{-\mu s}F(s)\bigg|_{s=0}^{s=t}-\int_0^ t(-\mu) e^{-\mu s} F(s)ds\\
    &=e^{-\mu t}F(t)-F(0)+\mu\int_0^ t \underbrace{e^{-\mu s}}_{0\leq ...\leq 1} \underbrace{F(s)}_{\geq 0}ds\\
    &\leq e^{-\mu t}F(t)+\mu t \sup_{0\leq s\leq t}F(s)\\
    &\overset!= (e^{-\mu t}+\mu t)F(t)\\
    &=(e^{-\mu t}+\mu t)\int_0^ tf(s)ds.
\end{align*}
In the second to last step (marked by $!$) we have used that $F$ is increasing. 
For $x\in[0,\frac12]$ we have $x\leq e^{-x}$. Indeed, $xe^x$ is increasing on $[0,\infty)$ and $\frac12 e^{\frac12}=\frac12\sqrt e\leq \frac12\sqrt 4=1$.  For $t\in[0,\frac1{2\mu}]$ we have $\mu t\in[0,\frac12]$ and the Lemma follows by estimating $\mu t\leq e^{-\mu t}$. 
\end{proof}
\begin{clem}\label{EstimateLem1_Supp}
Let $\epsilon>0$, $\omega\in\R$, $R\in C^0(\R)$ and $\xi$ solve $\dot\xi(t)-2\epsilon\sin^2(\omega t)\xi(t)=R(t)$. 
Then, for $t\in[0, \frac1{2\epsilon}]$
$$|\xi(t)|\leq |\xi(0)|e^{\frac\epsilon{2\omega}} e^{\epsilon t}+2e^{\frac\epsilon\omega} \int_0^t |R(s)|ds.$$
\end{clem}
\begin{proof}
Using $2\sin^2 x=1-\cos(2x)$, we compute
\begin{align}
    \frac d{dt}\left[\xi(t) e^{-\epsilon t+\frac{\epsilon\sin(2\omega t)}{2\omega}}\right]&=e^{-\epsilon t+\frac{\epsilon\sin(2\omega t)}{2\omega}}\left[\dot \xi(t) +\xi(t)\frac{d}{dt}\left(-\epsilon t+\frac{\epsilon\sin(2\omega t)}{2\omega}\right)\right]\nonumber\\
    &=e^{-\epsilon t+\frac{\epsilon\sin(2\omega t)}{2\omega}}\left[
    \dot \xi(t) +\xi(t)\left(-\epsilon+\epsilon\cos(2\omega t)\right)
    \right]\nonumber\\
    &=e^{-\epsilon t+\frac{\epsilon\sin(2\omega t)}{2\omega}}\left[
    \dot \xi(t)- 2\epsilon\sin^2(\omega t)\xi(t)
    \right]\nonumber\\
    &=e^{-\epsilon t+\frac{\epsilon\sin(2\omega t)}{2\omega}}R(t).\label{supp1neu}
\end{align}
Integrating gives 
$$\xi(t) e^{-\epsilon t+\frac{\epsilon\sin(2\omega t)}{2\omega}}-\xi(0)=\int_0^ t e^{-\epsilon s+\frac{\epsilon\sin(2\omega s)}{2\omega}}R(s)ds.$$
We now estimate 
\begin{align*}
    |\xi(t)|&\leq e^{\epsilon t-\frac{\epsilon\sin(2\omega t)}{2\omega}}|\xi(0)|+e^{\epsilon t-\frac{\epsilon\sin(2\omega t)}{2\omega}}\int_0^t e^{-\epsilon s+\frac{\epsilon\sin(2\omega s)}{2\omega}}|R(s)|ds\\
  &\leq e^{\epsilon t+\frac{\epsilon}{2\omega}}|\xi(0)|+e^{\epsilon t+\frac{\epsilon}{2\omega}}\int_0^t e^{-\epsilon s+\frac{\epsilon}{2\omega}}|R(s)|ds\\
     &\leq e^{\epsilon t+\frac{\epsilon}{2\omega}}|\xi(0)|+e^{\epsilon t+\frac{\epsilon}{\omega}}\int_0^t e^{-\epsilon s}|R(s)|ds.
\end{align*}
To estimate further we use Lemma \ref{estimateLem0_Supp} to estimate 
$$\int_0^t e^{-\epsilon s}|R(s)|ds\leq 2e^{-\epsilon t}\int_0 ^t |R(s)|ds\hspace{.5cm}\textrm{for $t\in[0,\frac1{2\epsilon}]$}.$$
Inserting this Estimate gives 
$$
|\xi(t)|\leq |\xi(0)|e^{\frac\epsilon{2\omega}}e^{\epsilon t}+2e^{\frac{\epsilon}{\omega}}e^{\epsilon t}e^{-\epsilon t}\int_0^ t|R(s)|ds
=
|\xi(0)|e^{\frac\epsilon{2\omega}}e^{\epsilon t}+2e^{\frac{\epsilon}{\omega}}\int_0^ t|R(s)|ds.
$$
\end{proof}

We now prove a criterion that ensures the convergence of the perturbation series in Equation \eqref{supppertuseries}.
\begin{clem}\label{suppconvergencelemma}
If $\delta >\epsilon$ and
$$24 e^{\frac{2\epsilon}\omega}|q_0|\left(|z(0)|+\frac1\delta\right)<1,$$
the series in Equation \eqref{supppertuseries} in convergent. 
\end{clem}
\begin{proof}
Put $t_0:=\frac1{2\delta}$ and $u_k:=\sup_{0\leq s\leq t_0}|z_k(s)|$. Applying Lemma \ref{EstimateLem1_Supp} to Equation \eqref{z0ODE_Supp} gives 
\begin{equation}\label{u0estimate_Supp}
u_0\leq |z(0)|e^{\frac\epsilon{2\omega}+\epsilon t_0}+2e^{\frac\epsilon{\omega}}t_0=:\alpha_0.
\end{equation}
For $n\geq 0$, applying Lemma \ref{EstimateLem1_Supp} to Equation \eqref{znODE_Supp} and subsequently using Lemma \ref{estimateLem0_Supp} gives
\begin{align}
    u_{n+1}&\leq 2 e^{\frac\epsilon\omega}|q_0|\sum_{j=0}^{n}\int_0^{t_0} e^{-\delta s} |z_j(s) z_{n-j}(s) |ds\nonumber\\
    &\leq 4 e^{\frac\epsilon\omega}|q_0|e^{-\delta t_0}\sum_{j=0}^{n}\int_0^{t_0}  |z_j(s) z_{n-j}(s) |ds\nonumber\\
    &\leq 4 e^{\frac\epsilon\omega}|q_0|e^{-\delta t_0}t_0\sum_{j=0}^{n}u_j u_{n-j}.\label{ukbound}
\end{align}
We use $x e^{-x}\leq 1$ for all $x\geq 0$ to estimate  $t_0e^{-\delta t_0}\leq\delta^{-1}$ and put $C:=4(e\delta)^{-1}e^{\frac\epsilon\omega}|q_0|$. For $n> 0$, we define $\alpha_n$ by 
\begin{equation}\label{alpharecursion_Supp}
    \alpha_{n+1}=C\sum_{j=0}^n\alpha_j\alpha_{n-j}.
\end{equation}
We claim $u_n\leq\alpha_n$ for all $n\geq 0$. For $n=0$ this is true by definition and for $n\geq 1$ it follows inductively. Indeed, assuming $u_k\leq \alpha_k$ for all $0\leq k\leq n$ we estimate 
$$u_{n+1}\overset{\eqref{ukbound}}\leq C\sum_{j=0}^n u_j u_{n-j}\leq C\sum_{j=0}^n \alpha_j \alpha_{n-j}\overset{\eqref{alpharecursion_Supp}}=\alpha_{n+1}.$$
Therefore $\sum_{n\geq 0} z_n\delta ^n$ converges absolutely when $\sum_{n\geq 0}\alpha_n\delta^n$ converges. To derive a criterion for the convergence of $\sum_{n\geq 0}\alpha_n \delta^n$ we consider the generating function $A(x):=\sum_{j\geq 0}\alpha_j x^j$. Using (\ref{alpharecursion_Supp}), we compute 
\begin{align*}
    Cx A^2(x)&=\sum_{n=0}^\infty Cx\left(\sum_{k=0}^\infty \alpha_k x^k\right)\left( \sum_{l=0} ^\infty \alpha_l x^l\right)\\
    &=Cx\sum_{n=0}^\infty \left[x^n \sum_{k=0}^n \alpha_k\alpha_{n-k}\right]\\
    &=\sum_{n=0}^\infty \left[x^{n+1} \left(C\sum_{k=0}^n \alpha_k\alpha_{n-k}\right)\right]\\
    &=\sum_{n=0}^\infty x^{n+1}\alpha_{n+1}\\
    &=A(x)-\alpha_0.
\end{align*}
This shows that $A(x)$ satisfies the quadratic equation $CxA(x)^2-A(x)+\alpha_0=0$. Therefore 
$$A(x)=\frac{1\pm\sqrt{1-4C\alpha_0 x}}{2Cx}.$$
The correct sign is $-$. Indeed, assume $+$ was correct. Then we get a contradiction as
$$\alpha_0=A(0)=\lim_{x\rightarrow 0} A(x)\overset!=\lim_{x\rightarrow 0}\frac{1+\sqrt{1-4C\alpha_0 x}}{2Cx}
\lim_{x\rightarrow 0}\frac2{2Cx}\hspace{.5cm}\textrm{, which is divergent}.$$
So 
\begin{equation}\label{Asolution_Supp} 
A(x)=\frac{1-\sqrt{1-4C\alpha_0 x}}{2Cx}.
\end{equation}
We use the expansion 
$$\sqrt{1-\epsilon}
=(1-\epsilon)^{\frac12}
=\sum_{n=0}^\infty \binom{\frac12}{n}(-\epsilon)^n
$$
to expand
\begin{align*} 
A(x)&=\frac1{2Cx}\left[1-\sum_{n=0}^\infty\binom{\frac12}n (-1)^n (4C\alpha_0x)^n\right]\\
&=-\frac1{2Cx}\sum_{n=1}^\infty\binom{\frac12}n (-1)^n (4C\alpha_0x)^n\\
&=\sum_{n=0}^\infty-\frac1{2C}\binom{\frac12}{n+1} (-4C\alpha_0)^{n+1} x^n.
\end{align*}
By definition $A(x)=\sum_{n\geq 0}\alpha_n x^n$. Comparing coefficients, we get 
$$\alpha_n=\frac{-1}{2C}\binom{\frac12}{n+1}(-4C\alpha_0)^{n+1}.$$
We use Stirling's approximation to get an asymptotic expansion of $\alpha_n$:
\begin{align*}
\binom{\frac12}{n}&=\binom{2n}{n}\frac{(-1)^{n+1}}{4^n(2n-1)}\\
&=\frac{(2n)!}{(n!)^2}\frac{(-1)^{n+1}}{4^n(2n-1)}\\
&\sim \frac{(2n)^{2n)}}{e^{2n}}\sqrt{4\pi n}\frac{(e^n)^2}{(n^n)^2(\sqrt{2\pi n}^2}\frac{(-1)^{n+1}}{4^n(2n)}\\
&=\frac{4^n}{\sqrt{\pi n}}\frac{(-1)^{n+1}}{4^n(2n)}\\
&=\frac{(-1)^{n+1}}{2\sqrt{\pi }}\frac1{n^{\frac32}}
\end{align*}
Using this asymptotic formula we get 
$$\alpha_n\sim \frac{-1}{2C}\frac{(-1)^n }{2\sqrt\pi}\frac1{\sqrt{(n+1)^{\frac32}}}(-1)^{n+1}(4C\alpha_0)^{n+1}=
\frac{1}{4C\sqrt\pi}\frac{(4C\alpha_0)^{n+1}}{(n+1)^{\frac32}}.$$
To ensure convergence of $\sum_{n\geq 0}\alpha_n \delta^n$ we must require 
$$1>\lim_{n\rightarrow\infty}\frac{\alpha_{n+1}\delta^{n+1}}{\alpha_n \delta^n}
=\delta \lim_{n\rightarrow\infty}\left[\frac{1}{4C\sqrt\pi}\frac{(4C\alpha_0)^{n+2}}{(n+2)^{\frac32}}\left(\frac{1}{4C\sqrt\pi}\frac{(4C\alpha_0)^{n+1}}{(n+1)^{\frac32}}\right)^{-1}\right]=(4C\alpha_0)\delta.$$
Inserting the definition of $C$ we get 
$$1\overset !>4C\alpha_0\delta 4\left(4(\delta e)^{-1}e^{\frac\epsilon\omega}|q_0|\right)\delta=16 e^{-1}\alpha_0 e^{\frac\epsilon\omega} |q_0|.$$
By definition $\alpha_0=u_0$. Hence 
$$16 e^{-1}\alpha_0 e^{\frac\epsilon\omega} |q_0|\leq 16 e^{-1} e^{\frac\epsilon\omega} |q_0|\left( |z(0)|e^{\frac\epsilon{2\omega}+\epsilon t_0}+2e^{\frac\epsilon{\omega}}t_0\right).$$
We have $t_0=\frac1{2\delta}\leq\frac1{2\epsilon}$. The last step is justified by requiring $\epsilon<\delta$. Hence 
\begin{align*}
    4C\alpha_0 \delta &\leq 16 e^{-1}\alpha_0 e^{\frac\epsilon\omega} |q_0|\\
    &\leq 16 e^{-1} e^{\frac\epsilon\omega} |q_0|\left( |z(0)|e^{\frac\epsilon{2\omega}+\epsilon t_0}+2e^{\frac\epsilon{\omega}}t_0\right)\\
    &\leq 16 e^{-1} e^{\frac\epsilon\omega} |q_0|\left( |z(0)|e^{\frac\epsilon{2\omega}+\frac12}+\frac4\delta e^{\frac\epsilon{\omega}}\right)\\
    &\leq 16 e^{-1} e^{\frac\epsilon\omega} |q_0|\left( |z(0)|e^{\frac\epsilon{2\omega}}\cdot 4+\frac4\delta e^{\frac\epsilon{\omega}}\right)\\
    &\leq 16 e^{-1} e^{\frac\epsilon\omega} |q_0|\left(4e^{\frac\epsilon\omega}\right)\left( |z(0)|+\frac1\delta e^{\frac\epsilon{\omega}}\right)\\
    &\leq 64e^{-1}e^{2\frac\epsilon\omega}|q_0|\left(|z(0)|+\frac1\delta\right)\\
    &\leq 24e^{2\frac\epsilon\omega}|q_0|\left(|z(0)|+\frac1\delta\right).
\end{align*}
In the last step we have used that $64e^{-1}=23.544...\leq 24$. So, if
$$24e^{2\frac\epsilon\omega}|q_0|\left(|z(0)|+\frac1\delta\right)<1,$$
the series $\sum_{n\geq 0}\alpha_n \delta^n$ converges and hence $\sum_{n\geq 0}z_n(t)\delta^n$ converges absolutely.  
\end{proof}
\ \\
\noindent
\textbf{Detailed proof of Corollary III.3}\ \\
\textbf{Step 1: A Zeroth Order Extraction Law}\ \\
Including only the first term of the perturbation series gives 
$$z(t)=p_0(t)+e^{-\epsilon t} p_1(t)$$
where $p_0$ and $p_1$ are $T$-periodic. We fix an arbitrary $t$ and put $p_0:=p_0(t)$, $p_1:=p_1(t)$ as well as $z_n:=z(t+nT)$. Then 
$$z_n=p_0+A^n p_1.$$
We consider the following two equations:
\begin{align*}
    &z_{n+1}-z_n =A^n (A-1) p_1\\
   & z_{n+2}-z_{n+1}=A^{n+1}(A-1)p_1.
\end{align*}
Subtracting $A$ times the first equation from the second gives
$$z_{n+2}-(1+A)z_{n+1}+Az_n=0.$$
Recalling that $y=1/z$, puttning $y_n:=y(t+nT)$ and multiplying by $y_ny_{n+1}y_{n+2}$, we get 
$$y_n y_{n+1}-(1+A) y_n y_{n+2}+A y_{n+1}y_{n+2}=0.$$
By definition $y=x-L-q=h-L$. Putting $h_n:=h(t+nT)$, we get
\begin{align*} 
0=&h_{n+2}h_{n+1}-(1+A)h_nh_{n+2}+Ah_{n+2}h_{n+2}\\
&+L^2-(1+A)L^2+AL^2\\
&-h_n L-Lh_{n+1}+(1+A)(Lh_{n+2}+h_n L)-A(h_{n+1}L+Lh_{n+2})\\
=&h_{n+2}h_{n+1}-(1+A)h_nh_{n+2}+Ah_{n+2}h_{n+2}\\
&+L(-h_n-h_{n+1}+h_{n+2}+h_n+Ah_{n+2}+Ah_n-Ah_{n+1}-Ah_{n+2})\\
=&h_{n+2}h_{n+1}-(1+A)h_nh_{n+2}+Ah_{n+2}h_{n+2}\\
&+L(-h_{n+1}+h_{n+2}+Ah_n-Ah_{n+1}).
\end{align*}
Rearranging, we get 
$$L=\frac{h_{n+2}h_{n+1}-(1+A)h_nh_{n+2}+Ah_{n+2}h_{n+2}}{-h_{n+2}+(1+A)h_{n+1}-Ah_n}.$$

\noindent
\textbf{Step 2: A First Order Extraction Law}\ \\
Including the first two terms of the perturbation series gives 
$$z(t)=p_0(t)+e^{\epsilon t} p_1(t)+e^{-\delta t}(p_2(t)+e^{\epsilon t} p_3(t)+e^{2\epsilon t} p_4(t))$$
where $p_k$ are $T$-periodic functions. Putting $z_n:=z(t+nT)$ and $p_k:=p_i(t)$ for $0\leq k\leq 4$, we get
$$z_n=p_0 +A^np_1 +B^n(p_2 +A^n p_3+A^{2n}p_4).$$
We now methodically combine these equations for various $n$ to get an identity with right hand side $0$. We begin by computing
$$z_{n+1}-z_n=A^np_1(A-1)+B^n(B-1) p_2 +B^n A^n (BA-1) p_3 +B^n A^{2n}(BA^2-1) p_4.$$
Hence 
\begin{align*}
    &z_{n+2}-z_{n+1}-A(z_{n+1}-z_n)\\
=&A^{n+1}p_1(A-1)+B^{n+1}(B-1) p_2 +B^{n+1} A^{n+1} (BA-1) p_3 +B^{n+1} A^{2(n+1)}(BA^2-1) p_4\\
&-\left(A^{n+1}p_1(A-1)+AB^n(B-1) p_2 +AB^n A^n (BA-1) p_3 +AB^n A^{2n}(BA^2-1) p_4\right)\\
=&B^n(B-1)(B-A)p_2
+B^n A^{n+1}(B-1)(BA-1)p_3
+B^n A^{2n+1}(BA-1)(BA^2-1)p_4.
\end{align*}
Next we compute 
\begin{align*}
    &z_{n+3}-z_{n+2}-A(z_{n+2}-z_{n+1})-B\left(
    z_{n+2}-z_{n+1}-A(z_{n+1}-z_n)\right)\\
    =&B^{n+1}(B-1)(B-A)p_2
    +B^{n+1} A^{n+2}(B-1)(BA-1)p_3
    +B^{n+1} A^{2n+3}(BA-1)(BA^2-1)p_4\\
    &-
    \left(
    BB^n(B-1)(B-A)p_2
+BB^n A^{n+1}(B-1)(BA-1)p_3
+BB^n A^{2n+1}(BA-1)(BA^2-1)p_4
    \right)\\
=&B^{n+1}A^{n+1}(A-1)(B-1)(BA-1)p_3+B^{n+1}A^{2n+1}(A^2-1)/BA-1)(BA^2-1)p_4.
\end{align*}
We simplify 
\begin{align*} 
&z_{n+3}-z_{n+2}-A(z_{n+2}-z_{n+1})-B\left(
    z_{n+2}-z_{n+1}-A(z_{n+1}-z_n)\right)\\
=&z_{n+3}-(1+A+B)z_{n+2}+z_{n+1}(A+B+AB)-ABz_n.
\end{align*}
So we get 
\begin{align*} 
&z_{n+3}-(1+A+B)z_{n+2}+z_{n+1}(A+B+AB)-ABz_n\\
=&
B^{n+1}A^{n+1}(A-1)(B-1)(BA-1)p_3+B^{n+1}A^{2n+1}(A^2-1)(BA-1)(BA^2-1)p_4.
\end{align*}
Now we compute 
\begin{align*} 
&z_{n+4}-(1+A+B)z_{n+3}+z_{n+2}(A+B+AB)-ABz_{n+1}\\
&-AB\left(z_{n+3}-(1+A+B)z_{n+2}+z_{n+1}(A+B+AB)-ABz_n\right)\\
=&B^{n+2}A^{2n+1}(A^2-1)(A^2-1)(BA-1)(BA^2-1)p_4.
\end{align*}
So, we arrive at the identity 
\begin{align*} 
&z_{n+5}-(1+A+B)z_{n+4}+z_{n+3}(A+B+AB)-ABz_{n+2}\\
&-AB\left(z_{n+4}-(1+A+B)z_{n+3}+z_{n+2}(A+B+AB)-ABz_{n+1}\right)\\
&-A^2B\left[z_{n+4}-(1+A+B)z_{n+3}+z_{n+2}(A+B+AB)-ABz_{n+1}\right.\\
&\left.-AB\left(z_{n+3}-(1+A+B)z_{n+2}+z_{n+1}(A+B+AB)-ABz_n\right)\right]\\
&=0.
\end{align*}
We collect terms 
\begin{align*}
    0=&+z_{n+5}\\
    &-z_{n+4}(1+A+B+AB+A^2B)\\
    &=+z_{n+3}(A+B+AB+AB(1+A+B)+A^2B(1+A+B)+A^3 B^2)\\
    &=-z_{n+2}(AB+AB(A+B+AB)+A^2B(A+B+AB)+A^3B^2(1+A+B))\\
    &=+z_{n+1}(A^2B^2+A^3B^2+A^3B^2(A+B+AB))\\
    &=-A^4B^3 z_n
\end{align*}.
We can simplify further by taking $n=0$ and combining terms:  
\begin{align*}
    0=&+z_{5}\\
    &-z_{4}(1+A+B(1+A+A^2))\\
    &=+z_{3}(A+B(1+A)A(1+A+A^2)+B^2A(1+A+A^2))\\
    &=-z_{2}(AB(1+A+A^2)+(A+1)B^2(A+A^2+A^3)+A^3B^3)\\
    &=+z_{1}(AB^2(A+A^2+A^3)+A^3B^3(1+A))\\
    &=-A^4B^3 z_0
\end{align*}

\renewcommand{\lstlistingname}{Mathematica Code}
\renewcommand{\lstlistlistingname}{Source Code}

\definecolor{codegreen}{rgb}{0,0.6,0}
\definecolor{codegray}{rgb}{0.5,0.5,0.5}
\definecolor{codeorange}{rgb}{1,0.49,0}
\definecolor{backcolour}{rgb}{.95,0.95,.96}

\lstdefinestyle{mystyle}{
    backgroundcolor=\color{backcolour},   
    commentstyle=\color{codegray},
    numberstyle=\tiny\color{codegray},
    stringstyle=\color{codeorange},
    basicstyle=\ttfamily\footnotesize,
    breakatwhitespace=false,         
    breaklines=true,                 
    captionpos=b,                    
    keepspaces=true,                 
    numbers=left,                    
    numbersep=5pt,                  
    showspaces=false,                
    showstringspaces=true,
    showtabs=false,                  
    tabsize=2,
    xleftmargin=10pt,
}

\lstset{style=mystyle}

\noindent
\addtocounter{section}{1}
\textbf{\large H. Simulations}
\noindent
\textbf{Simple model with/without noise}\ \\
We first define all parameters
\begin{lstlisting}[language=Mathematica, mathescape=true, breaklines=true, caption={Definitions}]
$\epsilon$= .01;
b = 2;
T = 3;
$\omega$ = 2 $\pi$/T;
$\theta$ = Exp[-$\epsilon$ b T];
\end{lstlisting}

Next, we define the noise function. To do so, we first choose a parameter $dt>0$ and define the function 
$$\operatorname{Bumb}(t)=\left\{
\begin{aligned}
1&\hspace{.5cm}\textrm{if $0\leq t\leq dt$}\\
0& \hspace{.5cm}\textrm{else}
\end{aligned}
\right.$$
We now generate a random sequence of numbers $R_i$ and define the noise function 
$$n(t):=\epsilon^2\sum_i R_i\operatorname{Bump}(t-i dt).$$

\begin{lstlisting}[language=Mathematica, mathescape=true, breaklines=true, caption={Definitions}]
dt = .5;
Bump[t_] = UnitStep[t] - UnitStep[t - dt];
TableR = RandomReal[{-1, 1}, 90/dt];
n[t_] = $\epsilon$^(2) Sum[TableR[[i]] Bump[t - i dt], {i, 1, 90/dt}];
\end{lstlisting}

Next, we implement the extremum seeking ODE
$$\dot y+\epsilon b (1-\cos(2\omega t))y+b y^2\sin(\omega t)=-b\epsilon^2\sin(\omega t)^3+n(t)\sin(\omega t).$$

To generate the graphics without noise the last term in this equation must simply be dropped. 

\begin{lstlisting}[language=Mathematica, mathescape=true, breaklines=true, caption={Definitions}]
s = NDSolve[{y'[t] + $\epsilon$ b (1 - Cos[2 $\omega$ t]) y[t] + 
      b y[t]^2 Sin[$\omega$ t] == -b $\epsilon^2$ Sin[$\omega$ t]^3 + 
      n[t] Sin[$\omega$ t], y[0] == 1.3}, y, {t, 0, 100}, 
   AccuracyGoal -> 15];
sol[t_] = y[t] /. s;

(*Plot*)
Plot[sol[t], {t, 0, 30}]
\end{lstlisting}

Next, we implement the extraction scheme 
$$L=\frac{(x_0-x_1)x_2+\theta (x_2-x_1)x_0}{x_0-(1+\theta) x_1+\theta x_2}.$$
This formula requires the knowledge of $\theta$. To extract $\theta$, we first define 
$$g:=\frac{(x_0-x_1)(x_2-x_3)}{(x_1-x_2)(x_0-x_3)}$$
and obtain $\theta$ as 
$$\theta=\frac{1-g}2-\frac1{2g}\sqrt{(1-g)^2-4g^2}.$$
We define extraction formulas for $L$ with the extracted and the exact value of $\theta$. The first is called $\operatorname{ExtraL}(t)$ and the second one $\operatorname{ExtraLCheat}(t)$.

\begin{lstlisting}[language=Mathematica, mathescape=true, breaklines=true, caption={Definitions}]
g[t_] = Min[
   1/3, ((sol[t] - sol[t + T]) (sol[t + 2 T] - 
        sol[t + 3 T]))/((sol[t + T] - sol[t + 2 T]) (sol[t] - 
        sol[t + 3 T]))];

Extra$\theta$[t_] = -(g[t] - 1)/(2 g[t]) - 
   1/(2 g[t]) Sqrt[(g[t] - 1)^2 - 4 g[t]^2];
       
ExtraL[
   t_] = ((sol[t][[1]] - sol[t + T][[1]]) sol[t + 2 T][[1]] + 
     Extra$\theta$[t] sol[t][[
       1]] (sol[t + 2 T][[1]] - sol[t + T][[1]]))/(sol[t][[
      1]] - (1 + Extra$\theta$[t]) sol[t + T][[1]] + 
     Extra$\theta$[t] sol[t + 2 T][[1]]);

ExtraLCheat[t_] = ((sol[t][[1]] - sol[t + T][[1]]) sol[t + 2 T][[
       1]] + $\theta$ sol[t][[
       1]] (sol[t + 2 T][[1]] - sol[t + T][[1]]))/(sol[t][[
      1]] - (1 + $\theta$) sol[t + T][[1]] + $\theta$ sol[t + 2 T][[
       1]]);
       
(*Plot for extracted $\theta$*)
Ptheta = Plot[{$\theta$, Extra$\theta$[t], $\theta$}, {t, 0, 30} , 
  PlotRange -> {{0, 30}, {.86, 1}}, 
  AxesLabel -> {t,  "$\theta$"}, LabelStyle -> {FontSize -> 15}, 
  PlotLegends -> {
    "Exact $\theta$", "Extracted $\theta$"}, 
  PlotStyle -> {RGBColor[0.368417, 0.506779, 0.709798], 
    RGBColor[0.880722, 0.611041, 0.142051], 
    RGBColor[0.368417, 0.506779, 0.709798]}]
    
(*Plot for $g$*)
Plot[g[t], {t, 0, 30}]

(*Plot for extracted $L$*)
PComp = Plot[{sol[t], ExtraL[t]}, {t, 0, 30}, PlotRange -> All, 
  AxesLabel -> {t, 
    "\!\(\*SubscriptBox[\(x\), \(Cl\)]\) vs \!\(\*SubscriptBox[\(x\), \
\(acc\)]\)"}, LabelStyle -> {FontSize -> 15}, AxesOrigin -> {0, -.2}, 
  PlotLegends -> {"\!\(\*SubscriptBox[\(x\), \(Cl\)]\)", 
    "\!\(\*SubscriptBox[\(x\), \(acc\)]\)"}]
\end{lstlisting}

The extraction scheme is also employed with an averaged value of $\theta$. First, we define an averaged value of $\theta$, that is obtained by averaging the extracted $\theta$ over intervals $[0, kT]$ for $k=1,2,3$. Afterwards, the extraction formula for $L$ is implemented using these averaged values.

\begin{lstlisting}[language=Mathematica, mathescape=true, breaklines=true, caption={Definitions}]
av$\theta$1 = 
 1/(T) NIntegrate[Extra$\theta$[s], {s, 0, T}, AccuracyGoal -> 5, 
   WorkingPrecision -> 10]
av$\theta$2 = 
 1/(2 T) NIntegrate[Extra$\theta$[s], {s, 0, 2 T}, AccuracyGoal -> 5, 
   WorkingPrecision -> 10]
av$\theta$3 = 
 1/(3 T) NIntegrate[Extra$\theta$[s], {s, 0, 3 T}, AccuracyGoal -> 5, 
   WorkingPrecision -> 10]
   
ExtraLnewmean1[
   t_] = ((sol[t][[1]] - sol[t + T][[1]]) sol[t + 2 T][[1]] + 
     av$\theta$1 sol[t][[
       1]] (sol[t + 2 T][[1]] - sol[t + T][[1]]))/(sol[t][[
      1]] - (1 + av$\theta$1 ) sol[t + T][[1]] + 
     av$\theta$1 sol[t + 2 T][[1]]);
ExtraLnewmean2[
   t_] = ((sol[t][[1]] - sol[t + T][[1]]) sol[t + 2 T][[1]] + 
     av$\theta$2 sol[t][[
       1]] (sol[t + 2 T][[1]] - sol[t + T][[1]]))/(sol[t][[
      1]] - (1 + av$\theta$2) sol[t + T][[1]] + 
     av$\theta$2 sol[t + 2 T][[1]]);
ExtraLnewmean3[
   t_] = ((sol[t][[1]] - sol[t + T][[1]]) sol[t + 2 T][[1]] + 
     av$\theta$3 sol[t][[
       1]] (sol[t + 2 T][[1]] - sol[t + T][[1]]))/(sol[t][[
      1]] - (1 + av$\theta$3) sol[t + T][[1]] + 
     av$\theta$3 sol[t + 2 T][[1]]);
     
(*Plot*)
Plot[{ExtraLnewmean1[t], ExtraLnewmean2[t], 
   ExtraLnewmean3[t]}, {t, 0, 30}, 
  PlotLegends -> {"L for \!\(\*SubscriptBox[\($\theta$\), \(1\)]\)", 
    "L for \!\(\*SubscriptBox[\($\theta$\), \(2\)]\)", 
    "L for \!\(\*SubscriptBox[\($\theta$\), \(3\)]\)"}, 
  AxesLabel -> {"t", "Extracted L"}, LabelStyle -> {FontSize -> 15}]
\end{lstlisting}

\noindent
\textbf{Including a Drift}\ \\
We first define all parameters
\begin{lstlisting}[language=Mathematica, mathescape=true, breaklines=true, caption={Definitions}]
T = 3;
$\omega$ = 2 $\pi$/T;
$\epsilon$ = .1;
$\delta$ = .4;
y0 = 2;
A = Exp[$\epsilon$ T];
q0 = .01;
A = Exp[$\epsilon$ T];
q[t_] = q0 Exp[-$\delta$ t];
\end{lstlisting}

Next, we define the noise in the same way as we did before
\begin{lstlisting}[language=Mathematica, mathescape=true, breaklines=true, caption={Definitions}]
dt = .5;
Bump[t_] = UnitStep[t] - UnitStep[t - dt];
TableR = RandomReal[{-1, 1}, 90/dt];
n[t_] = $\epsilon$^(2) Sum[TableR[[i]] Bump[t - i dt], {i, 1, 90/dt}];
\end{lstlisting}

We now implement the ODE
$$\dot y+2\epsilon\sin^2(\omega t)y+y^2\sin(\omega t)=-\epsilon^2\sin(\omega t)^3+\delta q(t)-\sin(\omega t) n(t).$$
To generate graphics without noise it again suffices to just drop the last term in this equation. 
We also define $h(t):=y(t)-q(t)$ and implement the extraction scheme 
$$L=\frac{h_{n+2}h_{n+1}-(1+A)h_nh_{n+2}+Ah_{n+2}h_{n+2}}{-h_{n+2}+(1+A)h_{n+1}-Ah_n}.$$
To generate graphics without noise it again suffices to just drop the last term in this equation. 

\begin{lstlisting}[language=Mathematica, mathescape=true, breaklines=true, caption={Definitions}]
s = NDSolve[{y'[t] + 2 $\epsilon$ Sin[$\omega$ t]^2 y[t] + 
      y[t]^2 Sin[$\omega$ t] == - $\epsilon$^2 Sin[$\omega$ t]^3 + \
$\delta$ q[t] - Sin[$\omega$ t] n[t], y[0] == y0}, y, {t, 0, 70}];
sol[t_] = y[t] /. s;
h[t_] = sol[t] - q[t];
L[t_] = (h[t + T] h[t] - (1 + A) h[t + 2 T] h[
       t] + A h[t + 2 T] h[t + T])/(-h[t + 2 T] + 
     h[t + T] (1 + A) - A h[t]);
 
(*Plot *)
Plot[{sol[t], L1[t]}, {t, 0, 30}, AxesOrigin -> {0, -.1}, 
 PlotRange -> {{0, 30}, {-.1, .6}}, LabelStyle -> {FontSize -> 20}, 
 AxesLabel -> {"t", 
   "\!\(\*SubscriptBox[\(x\), \(\(Cl\)\(\\\ \\\ \)\)]\)vs. \
\!\(\*SubscriptBox[\(x\), \(acc\)]\) "},
 PlotLegends -> {
   Row[{"Classical ES"}],
   Row[{"\!\(\*SubscriptBox[\(x\), \(acc\)]\) with noise"}]
   }]
\end{lstlisting}

Obtaining the graphics for various $\Gamma$ is achieved by generating plots for the various choices of parameters described in Subsection IV.C and combining the plots.

\end{document}